\documentclass[10pt,journal]{IEEEtran}
\usepackage{amsmath,amsfonts}
\usepackage{algorithmic}
\usepackage{algorithm}
\usepackage{array}
\usepackage[caption=false,font=normalsize,labelfont=sf,textfont=sf]{subfig}
\usepackage{textcomp}
\usepackage{stfloats}
\usepackage{url}
\usepackage{verbatim}
\usepackage{graphicx}
\usepackage{cite}
\begin{document}

\title{DRIVE-T: A Methodology for Discriminative and Representative Data Viz Item Selection for Literacy Construct and Assessment}

\author{Angela Locoro and Silvia Golia and Davide Falessi}
%\address[1]{Department of Economics and Management, University of Brescia, Italy (e-mail: {angela.locoro,silvia.golia}@unibs.it)}
%\address[2]{Department of Informatics Engineering, University of Roma Tor Vergata, Italy (e-mail: falessi@ing.uniroma2.it)}
%\tfootnote{``Research funded by the European Union – Next-Generation EU - PRIN 2022 D.D. 104 of 02-02-2022, Mission 4 Component 1 CUP D53D23008690006, project name: ``Characterizing and Measuring Visual Information Literacy'' ID 2022JJ3PA5.''}

\markboth{Journal of \LaTeX\ Class Files,~Vol.~14, No.~8, August~2021}%
{Shell \MakeLowercase{\textit{et al.}}: A Sample Article Using IEEEtran.cls for IEEE Journals}


\maketitle

\begin{abstract}
The underspecification of progressive levels of difficulty in measurement constructs design and assessment tests for data visualization literacy may hinder the expressivity of measurements in both test design and test reuse.  
To mitigate this problem, this paper proposes DRIVE-T (Discriminating and Representative Items for Validating Expressive Tests), a methodology designed to drive the construction and evaluation of assessment items. Given a data vizualization, DRIVE-T supports the identification of task-based items discriminability and representativeness for measuring levels of data visualization literacy. DRIVE-T consists of three steps: (1) tagging task-based items associated with a set of data vizualizations; (2) rating them by independent raters for their difficulty; (3) analysing raters' raw scores through a Many-Facet Rasch Measurement model. In this way, we can observe the emergence of difficulty levels of the measurement construct, derived from the discriminability and representativeness of task-based items for each data vizualization,  ordered into Many-Facets construct levels.
%(3) transforming the raw raters' scores into a logit score using Facets modelling, for the emergence and inspection of the resulting underlying construct, derived from the ordering of the difficulty of item-based tasks into levels. 
In this study, we show and apply each step of the methodology to an item bank, which models the difficulty levels of a measurement construct approximating a latent construct for data visualization literacy. This measurement construct is drawn from semiotics, i.e., based on the syntax, semantics and pragmatics knowledge that each data visualization may require to be mastered by people. %Each item was designed having in mind the  reference to one of these traits of the data visualization language. %the computation of the inter-rater agreement by several indices and aggregation facets, each capturing a nuance of the raters' concordance; 
The DRIVE-T methodology operationalises an inductive approach, observable in a post-design phase of the items preparation, for formative-style and practice-based measurement construct emergence. A pilot study with items selected through the application of DRIVE-T is also presented to test our approach. %Constructs-foregrounding is possible when the most discriminative and representative items are selected, which may help the community of test designers scrutinise, discuss and adjust items when they want to design or make sense of existing data visualization literacy assessment tests, whenever no data visualization literacy measurement construct was explicitly or completely outlined beforehand.
%(3) transforming the raw raters' scores into a logit score using Facets modelling, for the emergence and inspection of the resulting underlying construct, derived from the ordering of the difficulty of item-based tasks into levels. 
\end{abstract}


\section{Introduction and motivation}
\label{sec:intro} %for journal use above \firstsection{..} instead
Data Visualization Literacy (data viz literacy from now on) is a measurable property that has been defined in many ways, according to multi-faceted perspectives, and through several assessment tools~\cite{aoyama2003graph,Galesicetal2011,Merbitzetal1989,Lee2017551,krejci2020visual,locoro2021visual,yang2021explaining,camba2022identifying,firat2022p,CALVI,davis2024risks,Cui-Promises-Pitfalls}. In summary, data viz literacy should define people's capability to read and write data visualizations (data viz from now on) and, as a human capability, it should be conceptualised, operationalised, and measured by proper assessment tests.
Open issues largely debated in the research community related to how to accurately capture all of the aspects of data viz capability include whether it changes with time and whether there should exist core invariances\footnote{A recently accepted full paper at the next IEEE VIS 2025 venue witnesses of all these major concerns of the community (an unofficial pre-print is available at \url{https://osf.io/8e3ag})}. 
Indeed, earlier studies proposing data viz items explicitly relied on cognitive models of learning evolution. These models were, e.g., Bloom's taxonomy~\cite{bloom1956taxonomy}, Pinker's graph comprehension theory~\cite{pinker1990theory}, and the one by Curcio~\cite{friel2001making}. However, these works did not articulate a clear construct mapping between theoretical models and item characteristics, such as discriminability or representativeness~\cite{burns2020evaluate,boy2014principled}. Item discriminability is the ability of items to express progressive levels of data viz literacy. Item representativeness is the ability of items to represent each trait or each level of data viz literacy. %Although earlier approaches~\cite{burns2020evaluate,boy2014principled} may have resulted too abstract and rigid for successfully targeting the most dynamic learning aspects and evolutionary nature of data viz literacy,
In more recent data viz literacy stances, the need for a methodology for item preparation is a compelling requirement~\cite{Beschietal2025}. The main construct for item design in the above cases is mostly the definition of data viz literacy, which is usually made of one sentence, usually different for each research group (no standard still exists), and not detailed into a more precise model~\cite{Beschietal2025}. Literacy is rarely discussed in terms of levels of it, defining how levels are layered down in the latent construct, posing any initial hypothesis on what elements compose it, what should be measured, or what underlying assumptions should be made explicit about the target population, the measuring intent, and the assessment tools. Posing all of the above aspects with full certainty may lead to contentious implications, as designers are still committed to measurement hypotheses\footnote{See again the community stances available at \url{https://osf.io/8e3ag}}. The above depiction should not be misconstrued as evidence of poor practices within the data viz
literacy community. It is crucial to remark that it should rather represent cautious stances, adopted to avoid common biases when dealing with the problem of measuring a human's property~\cite{mari2023measurement}. Indeed, whatever results were obtained, their shared consensus would be at most inter-subjective, let alone completely \textit{ob-jective}\footnote{in subjective stances we usually concede that something will always remain under (sub-) defined, and beyond the control level.}.

One of the most complete, largely adopted and recent visualization literacy assessment tests, e.g., the Visualization Literacy Assessment Test (VLAT)~\cite{Lee2017551, Pandey20231}, does not rely on any explicitly described cognitive model. However, failing to hypothesize any explicit mapping between a measurement construct and dimensions expressed by items may compromise its validity, trust, and reuse, no matter how the underlying visualization literacy latent construct should remain underspecified, empirically grounded, or flexible enough to adapt to the complexity of real and evolving applications. Consequently, when designing the shortened versions of an assessment test, there is no justification for selecting items based on such a construct~\cite{Galesicetal2011,Okanetal2019}. 

Since a fully defined construct model such as that of Bloom can be defined as a full theoretical approach and VLAT can be defined as a full empirical approach, we argue that a third way would be preferable. Our proposal could alleviate the limitations of both, i.e., not making explicit the underlying assumptions~\cite{Cabitza2014MadeWK} of item design, nor adjusting them from being too rigid and narrow.

Investigating this phase of the design and proposing supporting approaches for better quality of items, transparency of the process and post-design analysis are necessary actions to be taken in order for data viz literacy measurement to gain momentum. Our study investigates the following research questions: 
\begin{enumerate}[RQ1]
    \item How to proceed inductively from a set of data viz items up to construct levels of data viz literacy?
\item To what extent is it possible to solve the construct blurriness issue at design time, i.e., without involving respondents?
\item How can we determine, during the design phase, whether data viz effectively discriminate among and represent distinct levels of data viz literacy? %(e.g., Can we derive from this an outline of a growing progression of data viz literacy levels)? %multifaceted aspects, such as the kind of data viz, the task at hand and / or the kind of item?
\end{enumerate}

We propose a methodology, whose steps fall within item modelling design~\cite{wilson2004constructing}, to support the revelation of measurement construct details, which are left mostly implicit in practice. %For these reasons, we intend to share light in the items preparation phase of a measurement activity.
DRIVE-T (Discriminating and Representative Items for Validating Expressive Tests) methodology is also intended to support the identification of tasks difficulty, discriminability and representativeness for measuring levels of data viz literacy. 

This paper describes a task-based item design effort as an applied example of our methodology. It shows the creation of a new assessment test for data viz literacy, and a Many-Facets Rasch Measurement (MFRM) model able to outline a measurement construct emerging from the rankingn of those items as difficulty levels of data viz literacy. To the best of our knowledge, the MFRM model has never been applied to a task of this kind, neither to the data viz domain nor to the problem of data viz literacy measurement. In our model, examinees are the data viz, and what is measured is their ability to discriminate and represent data viz literacy levels. This new passage and application of the model is properly formalised and justified in what follows, also in regard to the  above research questions.

We then validate the DRIVE-T ability to outline post-design progression levels of a measurement construct via a pilot study. After selecting a set of items from our item bank based on their discriminability and representativeness, we administered them to a sample of 72 high-school students. The aim of the pilot study was to show that the assessment test proposed was able to measure data viz literacy properly (with a minimum but complete set of task-based items per data viz). DRIVE-T is intended to be used with brand new items or to extend existing item banks. In the former case, it is advisable to follow up with a pilot study as detailed in Section 4.

The paper is structured as follows: Section~\ref{sec:approach} analyzes past work on task-based item design in relation to DRIVE-T; Section~\ref{sec:method} describes  DRIVE-T; Section~\ref{sec:results} reports the results of our methodology, and those of the pilot study; Section~\ref{sec:discussion} discusses the results of applying DRIVE-T to the design of an assessment test, of using a pilot study, and outlines the limitations of the study. Section~\ref{sec:conc} reports final considerations.

\section{Related Work}
\label{sec:approach}
This section discusses past studies on the generation of task-based item design. As a matter of fact, since the outset of human thoughts about knowledge, aspects of perception, cognition and learning were outlined in philosophical and scientific doctrines, from Aristotle to Peirce. Drawing from the Greek philosopher~\cite{aristotle2006nicomachean}, who characterised knowledge into epistemic (know what), technical (know how), and practical wisdom (know why), we can seamlessly cross the centuries until the semiotic turn~\cite{peirce1931semiotics}, one of the most universal and powerful explanations of how the representational, meaningful, and actionable aspects of language are capturing the essence of knowing. The semiotic paradigm is applicable to any system of signs, whether they are words, numbers or graphs, and it constitutes a proxy for investigating models of cognitive evolution. 

Data viz are commonly defined as visual interfaces, able to provide iconic representations of abstract concepts such as statistical constructs. A recent stance~\cite{10756169} outlines their status of being boundary objects, i.e., knowledge artifacts capable of crossing domains and communities with their universal language, able to vehiculate expert knowledge to laymen.  

Recently, De Souza~\cite{de2005semiotic} applied a semiotic engineering framework to interactive systems seen as meta-communicative artifacts, emphasizing that designers and users engage in an indirect yet profound communication through the system’s interface. According to De Souza, the interface embodies the designers' communicative intent, expressing their understanding of the users' needs, preferences, and contexts of use. By adopting this perspective, De Souza foregrounded the idea that effective interaction design is inherently about effectiveness in this communicative relationship, ensuring that users correctly interpret the designers' intended messages, which are encoded using syntactic, semantic, and pragmatic dimensions. %Furthermore, De Souza remarked the need of designers to deeply reflect on the meanings embedded in their interfaces, advocating that interfaces should explicitly facilitate user interpretation and sense-making. She argued that usability extends beyond mere functionality to include the clarity and coherence of the interface communicative content. Therefore, in semiotic engineering, evaluation of interactive systems explicitly involves assessing how clearly the designers' intentions are communicated and how easily users can interpret and apply these intentions to accomplish their tasks effectively. 
%This emphasis on meta-communication enhances both user empowerment and the overall quality of user experience, bridging cognitive, practical, and ethical dimensions of interaction into play.
% FINE SPOSTABILE / SACRIFICABILE

As outlined in the introduction, in data viz literacy modelling many theories were proposed, but none seemed to have survived nor have become the ground truth in the community~\cite{Beschietal2025}. 
The tasks by data type classification proposed by Schneidermann~\cite{shneiderman2003eyes}, who in his mantra emphasised interactional tasks such as ``overview, zoom, filter, and details-on-demand'', mainly addressed the direct manipulation of visual interfaces, without explicitly encompassing the deeper cognitive and communicative dimensions involved in data viz literacy. As anticipated in the Introduction, Bloom’s taxonomy, Curcio’s levels of visualization literacy, and Pinker’s model of graphical comprehension offer important cognitive insights, but were criticised for being too abstract and excessively rigid. Bloom’s taxonomy, while useful for educational objectives, presents cognitive processes in rigid hierarchical structures that overlook the dynamic and fluid nature of human cognition and learning. Curcio’s visualization literacy levels tend to compartmentalise visual understanding, inadequately addressing the holistic integration of cognitive and communicative skills needed for effective data viz literacy. Similarly, Pinker’s graphical comprehension model primarily addresses cognitive decoding of graphical elements but lacks a comprehensive integration of communicative and social value dimensions essential for evolving human interaction models.

In this work, we pose that data viz literacy may be the embroidery of visual and symbolic knowledge, educational background and practical experience. Semiotics is a fluid enough framework, where the mastery of one layer (e.g., syntactic) of a language of signs does not necessarily entail the mastery of a ``subsequent'' layer (e.g., semantics). This underspecification assumption and relative independence of levels makes semiotics a more robust, well-grounded and generalizable hypothesis to validate models of data viz literacy. For the above reasons, we argue that adopting a semiotic framework for tasks-based item design is a more effective attempt to disentangle the nature and development of cognitively robust yet flexible data viz interaction models, ultimately supporting a richer and more meaningful explanation of what it means to be data viz literate.

\begin{figure}[ht]
\centering
\includegraphics[width=.85\columnwidth]{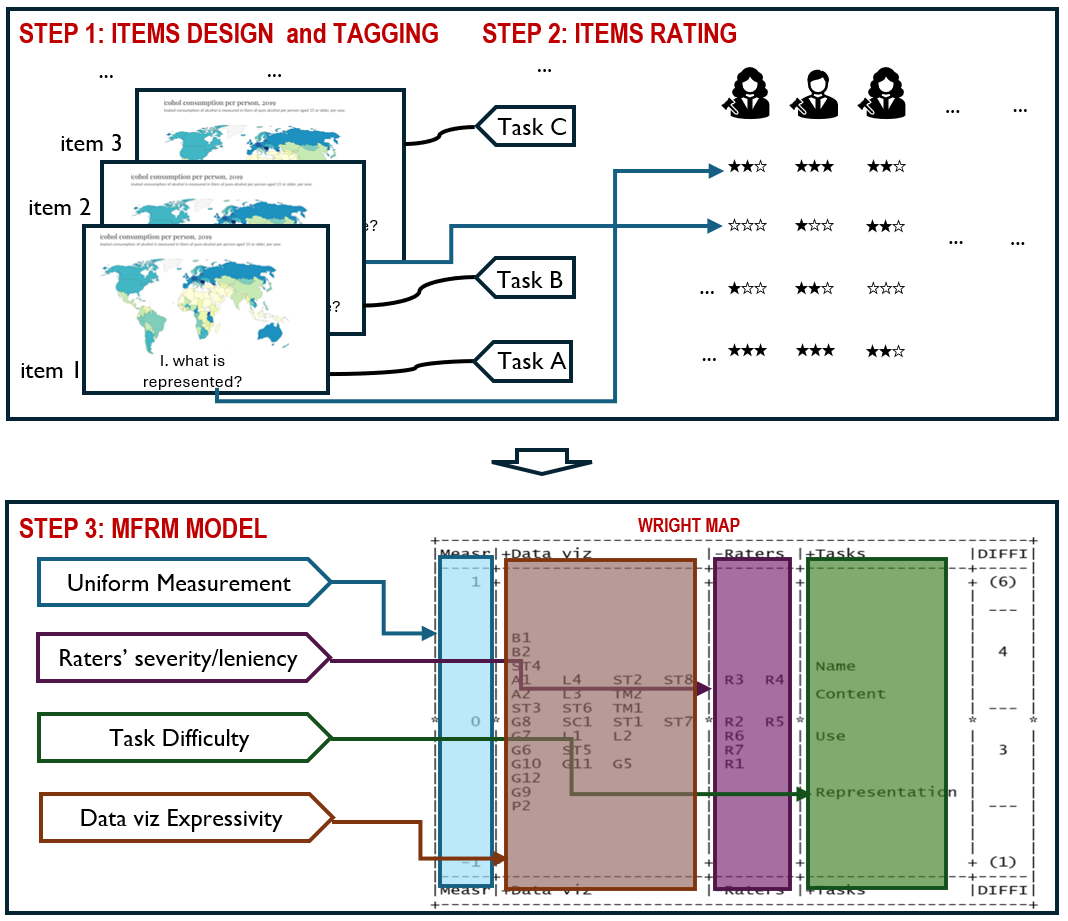}
\caption{The steps of the DRIVE-T methodology. Data Viz items may be either designed from scratch or extended from a previous item bank. Task-based tagging and raters' scoring are the two steps 1 and 2, while step 3 consists in the application of MFRM to model the task-based item scores by raters. This model allows to outline a ranking among raters and among tasks, each expressing: (i) data viz discriminability and representativeness; (ii) raters' severity; (iii) and tasks difficulty, respectively.}
\label{fig:flow}
\end{figure}

\section{Method}
\label{sec:method}
%%%%%%%%%%%%%%%%
%PARTE SPOSTATA QUI DALL'INTRO SU SUGGERIMENTO DI DAVIDE

This work tries to mitigate the problem of having a full measurement construct model of data viz literacy beforehand, in favor of a more ``formative style'', lazy-learning and inductive approach based on the design of new / tagging of existing items, which are supposed to measure different levels of the data viz literacy latent construct. In our view, %The intent is to foster the commonsense belief that %, when we design data viz literacy items, we have in mind some traits of growing difficulty that the items should discriminate among and be able to represent, as a proxy of this latent construct. In other words,
each item should challenge the respondent to express a capability related to its data viz literacy level. Current methodologies in social science propose to have a full construct hypothesis in mind beforehand~\cite{wilson2004constructing}, and to build items based on it. In the mingle of practice, these constraints are often blurred, relaxed, and less systematic than in theory. By considering the possibility of a more situated, still rigorous tool for item reliability checking, we may benefit the whole test-designers and test-takers community. Indeed, we would like to support the co-evolution of designing/refining/adjusting construct traits and items, and to help test-designers validate items/integrate explicit construct traits in their practice. In its turn, the validated construct should reflect and justify the choice of what data viz and items should compose a data viz literacy assessment test, according to the criteria of discriminability and representativeness. Discriminability refers to an item’s ability to distinguish among varying levels of data viz literacy in the target population, and representativeness determines what is the minimum set of items that should be selected among those available into an item bank, to uniquely but completely represent each measurable trait of the data viz literacy. Furthermore, test-designers of data viz literacy assessment tests should exploit this methodology to outline an on-the-fly hypothesis that may enhance their understanding of what and how many items should be used for measuring data viz literacy and, based on that, decide whether to refine/extend or keep the current items.

%%%%%%%%%%%%%%%
The DRIVE-T methodology is made of three steps, which are depicted in Figure~\ref{fig:flow}. Step 1 outlines a hypothesis of levels of data viz literacy borrowed from semiotics, and describes the process of tagging items according to semiotics tasks. Step 2 describes the process of rating them by judging/rating their difficulty. Step 3  supports the modelling of raters' scores with MFRM. Each step is detailed in the next subsections.

\subsection{Item design and tagging: Step 1}

%SPOSTABILE O SACRIFICABILE

Merging the two aspects of task-based literacy and semiotics, we propose a construct modelling activity characterised by task-based item design, where each task references one of the semiotic aspects of data viz language. Syntactic tasks are those that query the user about \textit{what a data viz represents} and strongly resonate with the structural, grammatical rule-based understanding of a formal system of signs, such as that of data viz. Semantic tasks have to do with the \textit{data viz content}, corresponding to a technical understanding of how to engage in a fruitful interaction with them. Such tasks extend beyond simple interaction-level manipulations to deeper interpretative actions, enabling users to extract informational gist. Pragmatic tasks embody reflective judgment in situated interactions. These tasks encourage users to actively consider the implications and appropriateness of \textit{a data viz to serve}, within a social context. Knowing \textit{a data viz name} may reveal a further, more profound level of knowledge, as knowing the name of an object is strongly associated with the ideal ``knowing it all'' about it\footnote{cf. The semiotician Umberto Eco in his thesis about the famous Gertrude Stein's verse ``A rose is a rose is a rose'', i.e., calling a thing with its name is showing meaning and knowledge of it.}.
%Such reflective engagement is highlighting the need for integrating pragmatic dimensions into interactive systems design to cultivate awareness and critical thinking skills.

\subsection{Item scoring: Step 2}
\label{sec:mirr}

Rater-mediated assessments are patently influencing critical decisions in educational, psychological, and hiring contexts, to name only a few. These assessments involve humans judging the performance, literacy, or other skills of the examinees. The problem with raters' evaluation in these kinds of tests is that it is fed with inevitably arbitrary behaviours such as interpretation, discretion, and subjectivity. Raters’ evaluations may vary due to differences in interpretation, severity or leniency biases,
and tendencies such as central tendency or halo effects biases~\cite{Eckes2015}. %More broadly, any measurement process (being it subjective or mediated by inanimate instruments) is hindered by its circular drift, often causing ``empirical adequacy of the theory or model and the reliability of measuring procedures [...] presuppose each other''~\cite[p.1160]{tal2013old}. %However, the solution to the measurement problem is not to avoid the raters' intervention nor to stop measuring. 

In the domain of assessment tests construction, raters may play a crucial role in acting as a self- or external validation device, able to provide a reliable pre-screening of how item design is gauged or needs to be adjusted. When not carefully managed, raters' variability can compromise the validity of this process. As much as rigorous is the raters' training, issues of the above kinds are not avoidable, hence a careful policy should be adopted in order to include raters' modelling in item design and assessment tasks.

Inter-rater (irr) reliability indices, available in common statistical packages, provide critical insights into rater agreement and consistency. %An overview and comparison of the main irr indices is reported in Table~\ref{tab:irrs}. 
However, these indices have some limitations and paradoxes~\cite{marasini2016assessing}. For instance, percentage agreement is among the most flexible and least stringent indices, merely reflecting the proportion of times raters agree exactly, but without correcting for chance agreement. In contrast, Fleiss' kappa is more stringent and less flexible, as it accounts for chance agreement and provides a more conservative estimate of reliability. However, collapsing some categories may bring a non-monotonic behaviour. Selecting an appropriate reliability index thus depends on the context and the level of precision required.
However, during item design or post-design analysis, there is the need to critically consider each item singularly. None of the methods above provide a fine-grained analysis of the raters' scoring able to shed light on their level of disagreements by individual items. For this reason, we are not going to carry out any irr computation and analysis. In this work, we apply an alternative analysis to the irr instead, by modelling the severity (resp. leniency) of raters with Rasch-based models. A brief presentation of individual items raw score differences and their visualization are proposed in Section~\ref{sec:results}. However, this score analysis is intended only to further comment on the main proposal of this work, i.e., providing a unified model of data viz, tasks, and raters. Indeed, some justifications of the model may be derived by a pinpoint identification of the controversies at item-level, where considerations may be drawn about the concurring aspects leading to disagreement (e.g., by comparing type, task-related activity or data viz choice). 

\subsection{Modelling raters' scores: Step 3}
\label{sec:familyRasch}
%%%%%%%%%%%%%%%%
%%%%%%%%%%%%%%%%
The family of Rasch models  consists of a set of probabilistic models that are able to convert raw scores into linear and reproducible measures~\cite{andrich1988rasch}. They require unidimensionality and local independence. The former refers to the requirement that all items within a questionnaire measure a single underlying construct, i.e., the latent trait of interest. The latter implies that, conditional on the latent trait, the responses to any given item are statistically independent of the responses to all other items in a questionnaire.
When data align with the model, the yielded measures are supposed to be objective. Being expressed in logit
(logarithm of odds), they are transformed into an interval measure. 

In the Rasch model (RM)~\cite{Rasch1960}, the probability that subject $n$ answers correctly ($c = 1$) or incorrectly
($c = 0$) to item $i$, depends on her/his level of latent trait (or ability), $\theta_n$, and on item difficulty, $\beta_i$, according to the following formula: 
\begin{equation}
P(X_{n}=c) = \frac{exp\left\{ c (\theta_n - \beta_i) \right\}}{1+exp\left\{\theta_n - \beta_i \right\} }\hspace{35pt} c = 0, 1
\label{eq:RM}
\end{equation}

The original RM was proposed to analyse tests composed exclusively of dichotomous items. Later, RM was extended to account for: (i) polytomous items, such as those on Likert scales, through the Rating Scale Model (RSM)~\cite{Andrich1978}, and the Partial Credit Model (PCM)~\cite{Masters1982}; and (ii) raters, who judged subjects (or examinees), through the Many-facet Rasch Measurement (MFRM) model~\cite{Linacre1989}.

In this paper, we will exploit the PCM and the MFRM model; the following subsections will provide their brief introduction.

%%%%%%%%%%%%%%%%
%%%%%%%%%%%%%%%%
\subsubsection{The Partial Credit Model}
\label{sec:PCM}
%%%%%%%%%%%%%%%%
%%%%%%%%%%%%%%%%
The PCM belongs to the family of RM that accounts for polytomously scored items. Given an item $i$ with $m+1$ response categories ($c = 0, 1, \cdots, m$), the probability of the subject $n$ to respond in category $c$ is given by:
\begin{equation}
P(X_{ni}=c) = \frac{exp\left\{ c (\theta_n - \beta_i) - \sum_{j=0}^c \tau_{ij} \right\}}{\sum_{l=0}^m exp\left\{l (\theta_n - \beta_i) - \sum_{j=0}^l \tau_{ij}) \right\} }
\label{eq:PCM}
\end{equation}
where $\theta_n$ denotes the level of latent trait (or ability) of subject $n$, $\beta_i$ represents the difficulty of item $i$ and $\tau_{ij}$, called thresholds, is the point of equal probability of categories $j-1$ and $j$ ($\tau_{i0} \equiv 0$ and $\sum_{j=1}^m \tau_{ij} =0$). The thresholds can be different for all the items.

Quality of PCM measures is obtained through the Person Reliability Index, the Outfit mean-square statistic (\textit{Outfit MNSQ}), and the Infit mean-square statistic (\textit{Infit MNSQ})~\cite{Linacre2002,MyfordWolfe2003}. 

The Person Reliability Index ranges from 0 to 1 and is robust to missing data. It provides an estimate of the replicability of one person's placement along the latent construct that can be expected if the same sample were administered a different set of items measuring the same underlying trait. 

\textit{Infit MNSQ} and \textit{Outfit MNSQ} assess the degree to which the observed data conform to the expectations of the Rasch analysis, facilitating the detection of aberrant individual responses or atypical response patterns at the item level. They are based on standardised residuals, range between 0 and infinity, and expected value of 1. Thus, values close to 1 reflect minimal distortion in the measurement system, values below 1 suggest overly predictable observations, while values above 1 indicate unpredictability and unmodeled noise in the data.

%%%%%%%%%%%%%%%%
%%%%%%%%%%%%%%%%
%\subsubsection{The Emergence of the construct levels: Step 2, Phase 1}
\subsubsection{The Many-facet Rasch Measurement Model}

\label{sec:MFRM}

The MFRM model can be seen as an extension of the basic RM, to include more variables (or facets), such as, for instance, raters evaluating the performance of examinees on tasks. A minimum of three facets configures the need to use a model of this kind~\cite{Eckes2009}. In this work, three facets were taken into account, i.e., examinees, raters and tasks. Thus, the MFRM model also accounts for raters' variability, i.e., the variability associated more with raters' characteristics than examinees' performance.

The MFRM model is formally expressed as:

\begin{equation}
\label{eq:MFRM}
\ln \left( \frac{p_{nijk}}{ p_{nijk-1}} \right) = \theta_n - \beta_i - \alpha_j   - \tau_k
\end{equation}

where $p_{nijk}$ is the probability that examinee $n$ receives a ranking of $k$ from rater $j$ on task $i$, $\theta_n$ is the proficiency of examinee $n$, $\beta_i$ is the difficulty of task $i$ and $\alpha_j$ is the severity of rater $j$. $\tau_k$ is called the threshold parameter and represents the difficulty of receiving a rating of $k$ relative to $k-1$ (thresholds add up to zero). The MFRM model defined in equation~\ref{eq:MFRM} involves three facets, and the scoring of the tasks shares the same rating scale structure with threshold parameters calibrated jointly across raters, tasks, and examinees. Thus, its name is Three-Facet Rating Scale Model (3FRSM)~\cite{LinacreWrite2002}.
In order to make the model identifiable, some constraints on its parameters must be imposed. In this work, the rater and task facets were centered, i.e., their mean values were constrained to be zero.

It is possible to use the separation reliability index (R) to evaluate the variability of the measures within each facet. This index describes how effectively the elements within the facet are separated, ensuring that the facet can be reliably defined. 
R is computed as the proportion of the observed variance of the facet measure which is not due to measurement error, and assumes values between 0 and 1. The interpretation of R depends on the facet considered. At the examinees level, R measures how well the model is able to differentiate them based on their proficiency, 
thus a high value is desirable. At the item level, R provides insights into how tasks vary in difficulty, being high values ideal in that they better encompass a broader spectrum of performance features, distributed across the underlying difficulty dimension. Finally, at the raters (judges) level, R provides information on the degree of similarity or dissimilarity between their severity/leniency. Thus, a low value of the R index is preferable whenever the raters are expected to be nearly interchangeable. 

%To investigate the accuracy of the examinees, raters, or tasks measurements, two fit statistics were exploited, based on standardized residuals: outfit mean-square statistic (\textit{Outfit MNSQ}) and infit mean-square statistic (\textit{Infit MNSQ})~\cite{Linacre2002,MyfordWolfe2003}. The standardized residual for examinee $n$, rated by the rater $j$ on task $i$ is given by the difference between the score obtained by examinee $n$ on task $i$ from the rater $j$ and the expected rating based on the estimates of the MFRM parameters, over the variance of the model.
%Infit and Outfit MNSQ have both expected value equal to 1, and they range between 0 and infinity. Therefore, values close to 1 indicate minimal distortion in the measurement system, values below 1 suggest overly predictable observations, while values above 1 indicate unpredictability and unmodeled noise in the data.

To investigate the accuracy of the examinees, raters, or tasks measurements, one can use the \textit{Outfit MNSQ} and \textit{Infit MNSQ} described in section~\ref{sec:PCM}. 
There is also another indicator that can be used to evaluate the examinee measurement, namely the point-measure correlation (\textit{PtMea Corr}): the Pearson correlation between the observed scores and the combined measures. The combined measure for examinee $n$, rated by the rater $j$ on task $i$ is given by
$\hat{\omega}_{nij}=\hat{\theta}_n - \hat{\beta}_i -\hat{\alpha}_j$, 
where $\hat{\theta}_n$, $\hat{\beta}_i$ and $\hat{\alpha}_j$ are the estimated parameters of equation~\ref{eq:MFRM}. \textit{PtMea Corr} offers insights into how observations align with the model expectations. When an examinee receives scores almost in line with the combined measures, the \textit{PtMea Corr} is positive and high, whereas in case of strong departure from model expectations, this index may assume negative values~\cite{Eckes2015}.

The model defined in formula~\ref{eq:MFRM} does not account for interactions between facets. However, studying the examinee-by-rate interactions may be informative with respect to the fairness of the judgment process. Thus, an interaction study may be performed by estimating the following model, which includes an examinee-by-rate interaction parameter $\phi_{nj}$, also known as a bias parameter: 

\begin{equation}
\label{eq:MFRM_int}
\ln \left( \frac{p_{nijk}}{ p_{nijk-1}} \right) = \theta_n - \beta_i - \alpha_j - \phi_{nj}  - \tau_k.
\end{equation}

Evaluating the significance of the examinee-by-rate interaction parameter $\phi_{nj}$ is possible by using a bias statistic $t_{nj}=\hat{\phi}_{nj}/SE_{nj}$, where $\hat{\phi}_{nj}$ was the estimated bias parameter and $SE_{nj}$ its standard error. This statistic is approximately distributed as a student $t$, under the null hypothesis that there is no bias apart from measurement error~\cite{Eckes2015}.

%%%%%%%%%%%%%%%%
\begin{figure}[ht]
\centering
\includegraphics[width=.85\columnwidth]{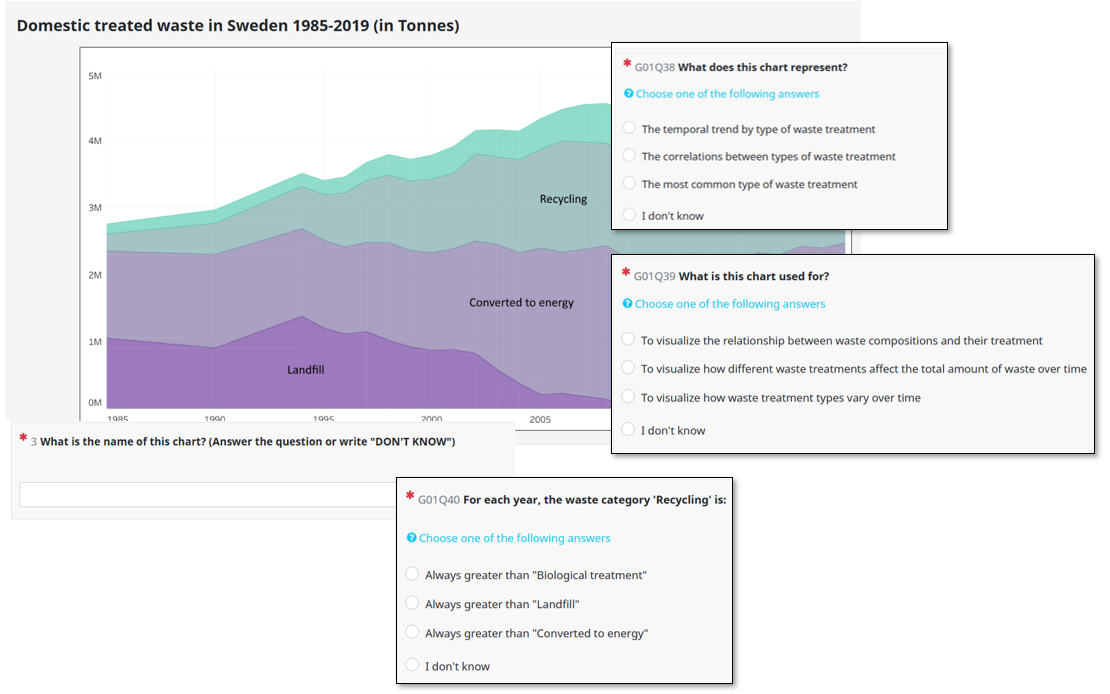}
\caption{Example of 4 task-based items designed for one of the 8 data viz included in our 44 task-based item bank. Each item in the figure corresponds to a task among: Name (bottom left item), Content (bottom right item), Represent (upper right item), and Use (rightward item).}
\label{fig:items}
\end{figure}
%%

%%%%%%%%%%%%%%%%
%%%%%%%%%%%%%%%%
\section{Results}
\label{sec:results}
%%%%%%%%%%%%%%%%
%%%%%%%%%%%%%%%%
%%%%%%%%%%%%%%%%
\subsection{Applying the DRIVE-T methodology to our item bank}
\label{sec:ibank}
%%%%%%%%%%%%%%%%
%%%%%%%%%%%%%%%%

%Data Visualization Literacy (DVL) 

In Step 1 of our visualization literacy assessment test, we decided to choose a subset of data viz related to the state-of-the-art assessment tests (see for example the MINI-VLAT test~\cite{Pandey20231}): a stacked area chart (A), a bubble chart (B), a choropleth map (G), a line chart (L), a pie chart (P), a scatter plot (SC), a stacked bar chart (ST) and a tree map (TM).
As pointed out in~\cite{Beschietal2025}, most papers proposing a questionnaire to measure data viz literacy use tasks which require ``purely visual operations or mental projections on a graphical representation''~\cite{boy2014principled}, such as, for example, identifying a maximum, a minimum, or a quantitative variation. 
As introduced in Section~\ref{sec:method}, we adopted a design approach with different aspects of comprehension in mind. These aspects are: the ability to read the data viz and understand what it represents (i.e., syntax), the ability to extract information from it (i.e., semantics), the knowledge of how it is used for serving its purpose (i.e., pragmatics), and knowing its name.
We labeled the corresponding tasks as \textit{Represent}, \textit{Content}, \textit{Use}, and \textit{Name}, respectively.

The formulation of each task was data viz specific. In the production phase, we designed 44 task-based items for most of the data viz, as summarised in table~\ref{tab:ntask}. An example of items, one for each task type, is reported in Figure~\ref{fig:items}\footnote{The items are available at \url{https://osf.io/ngu5q/?view_only=9f4144dcec4e49dabbebc332f539b3c8}. The items are in Italian despite an English informal  translation is provided as well).}.

\begin{table}[htb]
  \caption{Distribution of item versions by task type and associated data viz format.}
  \label{tab:ntask}
  \scriptsize
  \centering%
  \begin{tabular}{lcccc}
    \hline
    Data Viz &Name &  Represent & Use & Content \\
  \hline
    A & 1 & 1  & 1 & 2 \\
    B & 1 &  1 & 1 & 2 \\
    G & 1 & 2  & 2 & 2 \\
    L & 1 & 2 &  1 &  2  \\
    P & 1 & 1 & 1  &  2 \\
    SC& 1 & 1 & 1  & 1 \\
    ST& 1 & 2 & 2  & 2 \\
    TM & 1 & 1 &  1 & 2 \\
  	\hline
  \end{tabular}%
\end{table}

Multiple-choice items with three possible answers (and only one correct) were 54.4\% (24 items). Polytomous items on a 3-point Likert scale were 13.6\% (6 items). True/false were 11.4\% (5 items). One item had a free text answer. The last 8 items (20.6\%) were those connected with the task \textit{Name}, and they were free text. All items included the \textit{I don't know} option to prevent guessing.

To evaluate the quality of the items and decide which should compose the final questionnaire, we conducted a performance evaluation of the items and data viz, involving seven expert raters (judges). Raters were asked to assess each of the 44 items based on a six-point Likert scale of difficulty, from very easy to very difficult, shown in Table~\ref{tab:score}. To facilitate the rating activity and reduce discrepancies among raters, we defined a set of probability intervals for the probability of correct response, as shown in the table. 
\begin{table}[htb]
  \scriptsize
  \caption{Category labels, scores and probabilities of right response in percentage.}
  
  \label{tab:score}
  \centering%
  \begin{tabular}{l|r|r}
\hline
    Category label & Score &  Probability \%  \\
    \hline
    Very Easy & 1 &  95-100 \\
    Fairly Easy & 2 & 75-94 \\
    Middle Easy & 3 & 50-74 \\
    Middle Difficult &4 & 30-49 \\
    Fairly Difficult& 5 &6-29 \\
    Very Difficult& 6 &  0-5\\
  	\hline
  \end{tabular}
\end{table}

Each rater was asked to score the difficulty of each item without comparing it to the other items and to each other, as the focus was on determining the level of difficulty of each individual item on the provided scale. %(and based on the percentage of respondent who were supposed to give the correct answer). %for a examinee to provide the correct answer to each individual item. 
At the end of this procedure, we obtained a set of 44 evaluations for each rater.

%%%%%%%%%%%%%%%%
\begin{figure}[ht]
\centering
\includegraphics[width=.90\columnwidth]{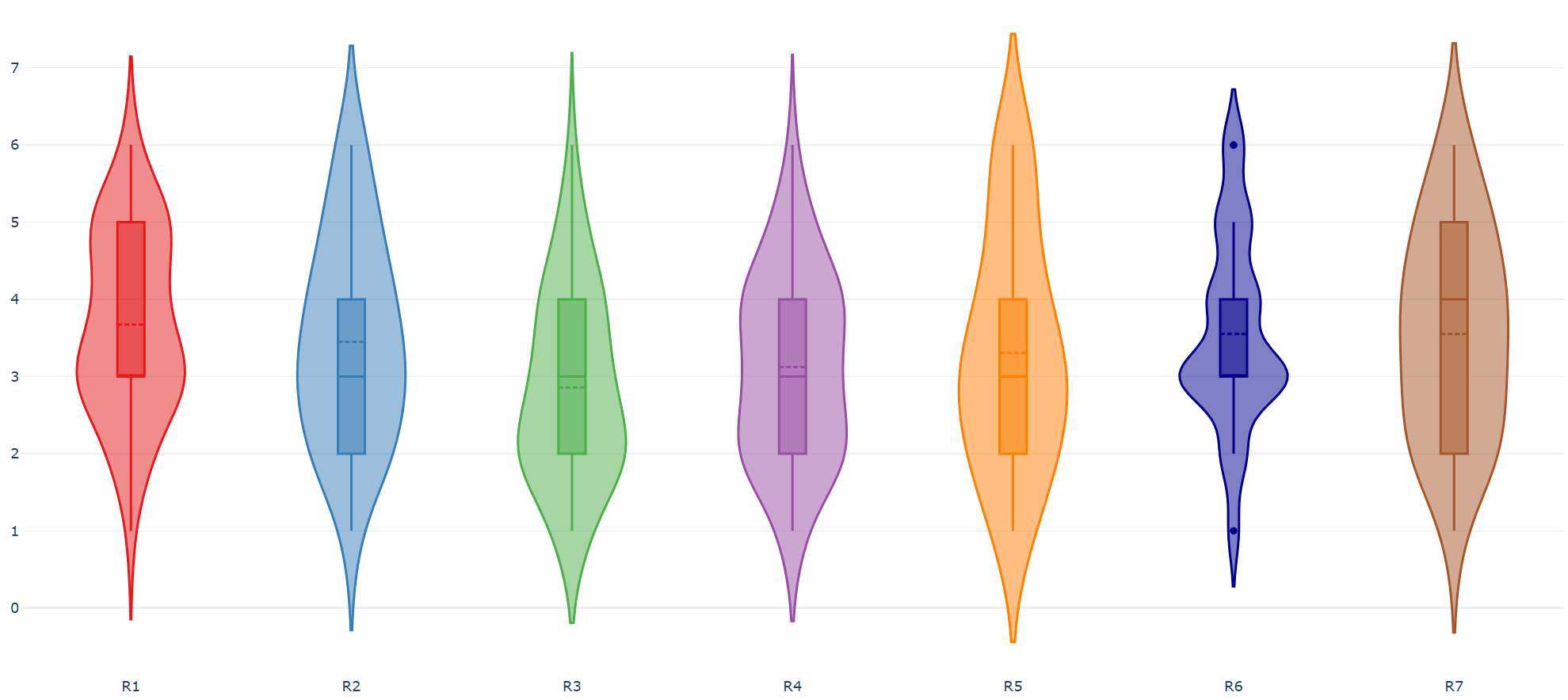}
\caption{Violin plots of the raters' scoring distributions.}
\label{fig:violin}
\end{figure}

\subsection{The raters' score}
\label{sec:irr}
In this study, the seven raters were recruited among scholars of two universities and among external advisors, with different expertise in data visualization. Two raters are in the research domain of data visualization (R3 and R4), three are in the research domain of statistics (R5, R6, and R7), and two are in the domain of informatics engineering and machine learning (R1 and R2). 
The single performance of raters concerning their use of the scale and the distribution of scores along the range of values is depicted in the violin plots of Figure~\ref{fig:violin}. Violin plots overlap to a more common box plot a distribution density that may help identify nuances in the use of scores, without looking at the $y$ axes.

%%%%%%%%%%%%%%%%
\begin{figure}[ht]
\centering
\includegraphics[width=\columnwidth]{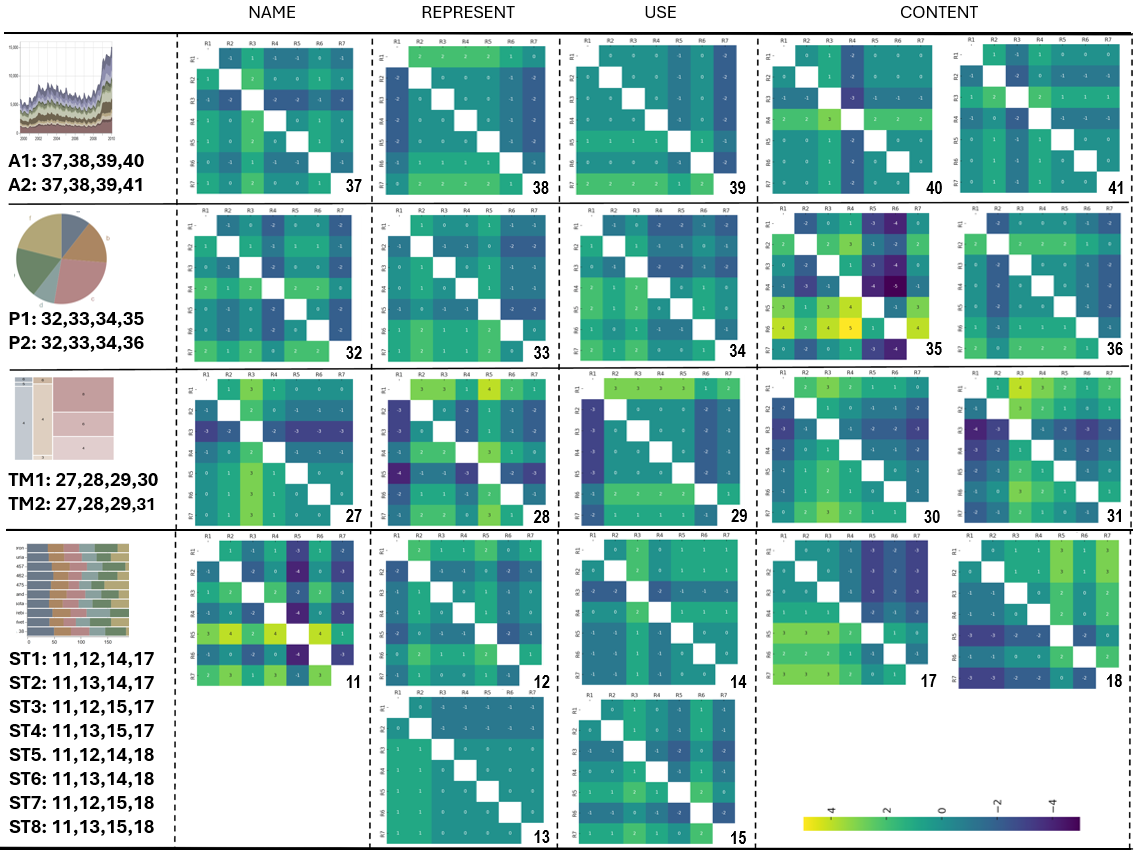}
\caption{Heatmaps comparisons among data viz task-based items by raters' ordered pairs. Each heatmap shows: at the top left rater R1 and at the bottom left rater R7, as per row disposition; at the top left rater R1 and at the top right rater R7, as per columns disposition. The data viz presented in the figure serve illustrative purpose only; they are of the same kind of those employed in the real tests: A, P, TM and ST. The item ID is the number at the bottom right of each heatmap, codified in Table~\ref{tab:comb}.}
\label{fig:heatmaps1}
\end{figure}

\begin{figure}[ht]
\centering
\includegraphics[width=\columnwidth]{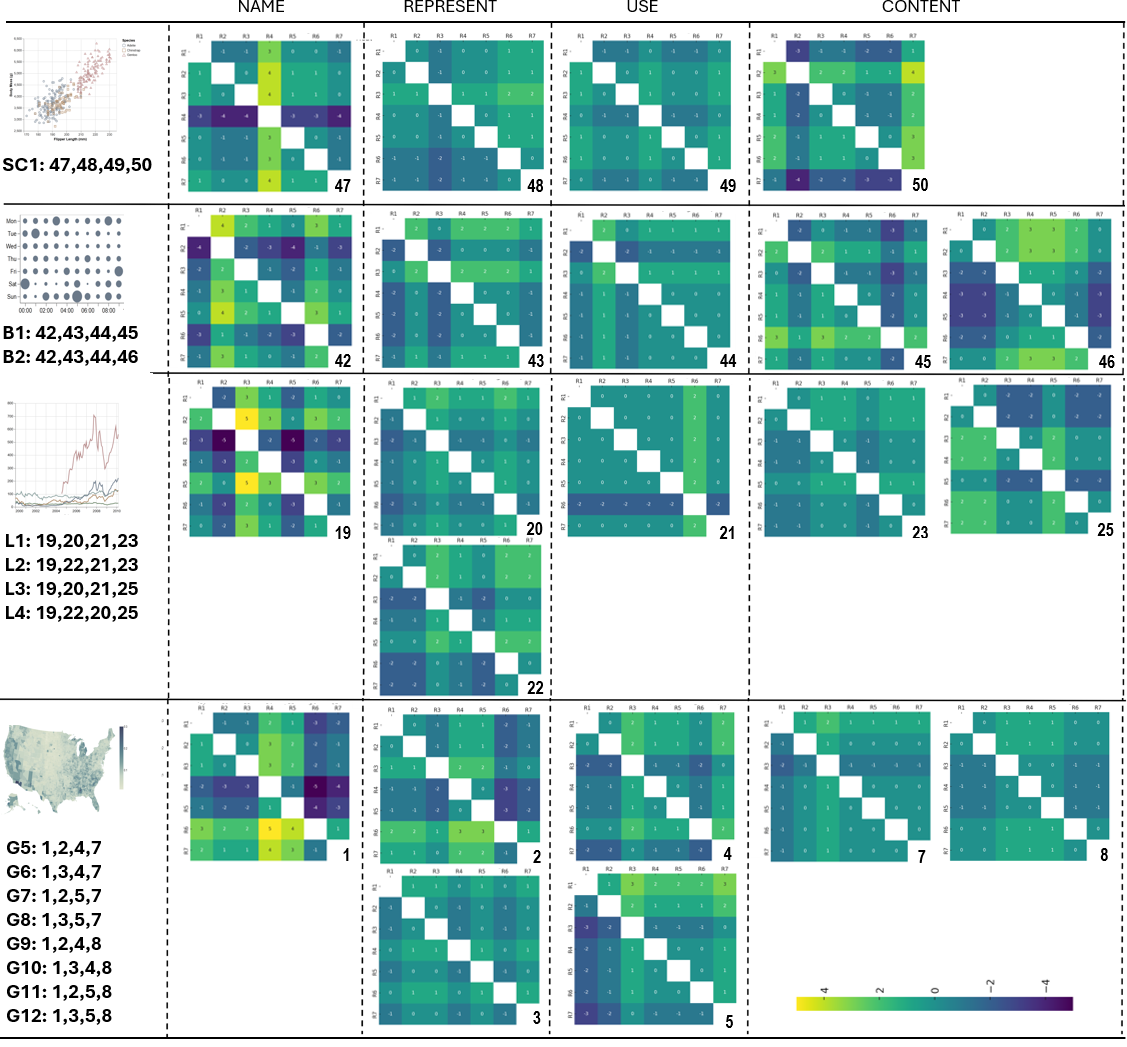}
\caption{Heatmaps comparisons among data viz task-based items by raters' pairs (same arrangement as Figure~\ref{fig:heatmaps1}). The data viz presented in the figure serve illustrative purpose only; they are of the same kind of those employed in the real tests: SC, B, L and G.}
\label{fig:heatmaps2}
\end{figure}

As anticipated in Section~\ref{sec:mirr}, no irr analysis was conducted on the raters' scores, but an item level analysis of raters' scores absolute differences. The results are reported in Figures~\ref{fig:heatmaps1} and~\ref{fig:heatmaps2}, where each heatmap is showing the score difference between ordered pair of raters. The higher the difference the lighter/darker the color of each cell of the heatmap (yellow for max positive differences and dark blue for max negative differences). Considering a difference of agreement of $0$ as perfect agreement, of $\pm{1}$ as very low disagreement, of $\pm{2}$ as low disagreement, of $\pm{3}$ as medium disagreement, and $\pm{4}$ or $\pm{5}$ as high disagreement, the analysis of the score differences among raters also reveals the degree of disagreement, whilst the sign of the difference allows to compare the severity / leniency of raters for each item. In particular, when the pairwise comparison gives a negative score difference, a darker green-to-blue color, this means that the row raters were more severe than the column raters (i.e., the row rater gave a lower score than the column rater); on the contrary, a positive score difference, a lighter green-to-yellow color, means that row raters were more lenient than column raters (i.e., the row rater gave a higher rating to the item, considering it more difficult).

%%%%%%%%%%%%%%%%
%%
%%

%%%%%%%%%%%%%%%%
%%%%%%%%%%%%%%%%
\subsubsection{The Many-Facet Rasch Measurement model of our item bank}
\label{sec:mfrmres}
%%%%%%%%%%%%%%%%
%%%%%%%%%%%%%%%%
In this section, we present the results obtained by estimating the 3FRSM, as defined in Equation~\ref{eq:MFRM}, using the joint maximum likelihood estimation method\footnote{This analysis was performed using the FACET software~\cite{FACET}, a common package mostly used in RM analyses.}. 
%

%, rather than a data viz judged as easy. In the latter case, most people will be able to read and retrieve correct information from the data viz, thus nearly everyone will respond correctly to questions related to it.

We distilled, from the initial banks of 44 items, 4 task-based items for each of the 8 data viz. We created a set of 29 records, each consisting of the combination of one data viz, one item for the \textit{Name} task, one item for the \textit{Represent} task, one item for the \textit{Use} task, and one item for the \textit{Content} task. %An anticipation of this codification was given when showing the raters' heatmaps (Figures~\ref{fig:heatmaps1} and~\ref{fig:heatmaps2}).

Hereafter, we use the term combination for the byproduct of a data viz and its four tasks, and identify each combination with the Label column of Table~\ref{tab:comb} (e.g., A1, A2, and the like). Table~\ref{tab:comb} also reports, on each line, the item number of each task. These 29 labels identify the 29 examinees (data viz combination) of our analysis, rated by the seven raters (judges) on the four tasks.

\begin{table}[ht]
  \scriptsize
  \caption{Combinations of data viz, \textit{Name}, \textit{Represent}, \textit{Use} and \textit{Content} tasks. The numbers refer to the numbers associated with each item.}
  \label{tab:comb}
  \centering%
  \begin{tabular}{llcccc}
    \hline
    Data Viz & Label &Name &  Represent & Use & Content \\
    \hline
    A & A1 & 37 & 38 & 39 & 40 \\
    & A2 & 37 & 38 & 39 & 41 \\
    B & B1 & 42  & 43 & 44  & 45\\
    & B2 & 42  & 43 & 44  & 46\\
    G & G5  & 1 & 2 & 4 & 7\\
    & G6  & 1 & 3 & 4 & 7\\
    & G7  & 1 & 2 & 5 & 7\\
    & G8  & 1 & 3 & 5 & 7\\
    & G9  & 1 & 2 & 4 & 8\\
    & G10  & 1 & 3 & 4 & 8\\
    & G11  & 1 & 2 & 5 & 8\\
    & G12  & 1 & 3 & 5 & 8\\
    L & L1  & 19 & 20 & 21 & 23\\
    & L2  & 19 & 22 & 21 & 23\\
    & L3  & 19 & 20 & 21 & 25\\
    & L4  & 19 & 22 & 21 & 25\\
    P & P1  & 32 & 34 & 33  & 35\\
    & P2  & 32 & 34 & 33  & 36\\
    SC& SC1  & 47 & 48 & 49 & 50\\
    ST & ST1  & 11  & 12 & 14 & 17\\
    & ST2  & 11  & 13 & 14 & 17\\
    & ST3  & 11  & 12 & 15 & 17\\
    & ST4  & 11  & 13 & 15 & 17\\
    & ST5  & 11  & 12 & 14 & 18\\
    & ST6  & 11  & 13 & 14 & 18\\
    & ST7  & 11  & 12 & 15 & 18\\
    & ST8  & 11  & 13 & 15 & 18\\
    TM & TM1  & 27 & 28 & 29 & 30\\
     & TM2  & 27 & 28 & 29 & 31\\
  	\hline
  \end{tabular}%
\end{table}

As anticipated in Section 1, in this study the role of the examinees was played by the data viz combinations. %What should be measured was the discriminability power, i.e., the ability to discriminate the level of data viz literacy owned by the test takers, as well as the representativeness, i.e., their ability to represent each of the tasks related to the semiotics traits of a measurement construct.
The hypothesis is that, given a data viz combination:

\begin{itemize}
\item the more it is judged to be difficult, the higher its discriminative power in assessing a person's data viz literacy level
\item the more it is complete and minimal (only one item per task, all the tasks being considered), the more representative it is
\end{itemize}

\begin{figure}[ht]
\centering
\includegraphics[width=.85\columnwidth]{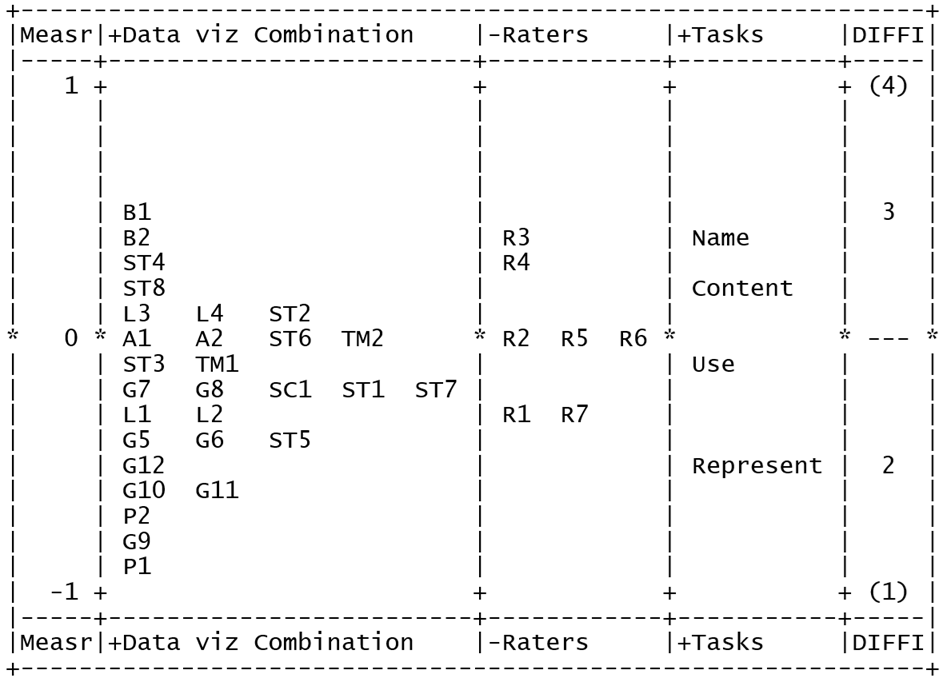}
\caption{Wright map}
\label{fig:WM}
\end{figure}

In our model, the data viz combinations were measured positively, i.e., high raw scores meant a high degree of discriminability.
The tasks were measured positively too, i.e., high difficulty measures corresponded to high scores. This result implies that, in the formula~\ref{eq:MFRM}, the item term was added instead of subtracted. 
%On the contrary, the raters were measured negatively, i.e., high severity measures of raters were indicated by low scores awarded to the data viz.
In contrast, raters were modeled such that higher severity corresponded to lower scores assigned to data visualizations.

The estimates of the threshold parameters $\tau_k$ resulted as in the following: $\tau_1$ = -1.54, $\tau_2$ = -0.33, $\tau_3$ = 0.16, $\tau_4$ = 0.88, $\tau_5$ = 0.83. The last two thresholds were very close and inverted, causing a disorder in the thresholds progression, which no longer advanced monotonically with the response categories (the ratings). 
Even if the requirement of the ordered thresholds is not part of the model constraints, we decided to overcome the problem by collapsing the first two and the last two categories, yielding a four-point Likert scale for rating difficulties. The resulting new thresholds resulted as no longer disordered.

A summary of the expected outcome statistics of the 3MFRM analysis is reported in Table~\ref{tab:mainq} of the Appendix~\ref{sec:appendix}.

The data viz, raters and tasks facets can be displayed in a single visualization called \textit{Wright map} (Figure~\ref{fig:WM}), as long as all three measurements are expressed in logit~\cite{Eckes2015}. The \textit{Measr} column displays the measurement scale. The \textit{Data viz Combination} column displays the estimated measure of discrimination ability, arranged with those at the top having the highest scores. The \textit{Raters} column displays the estimated measure of the raters' level of severity. The \textit{Tasks} column displays the average measure of difficulty of the four tasks considered; tasks listed higher were the more difficult. 

\subsection{The pilot study}
\label{sec:pilot}
One of the final outcomes of the DRIVE-T methodology was a more justified and robust selection of a subsample of items from our item bank, based on the rating activity of a group of raters. The analysis of the data performed by estimating the 3FRSM, came up with the Write map of Figure~\ref{fig:WM}. From the inspection of the \textit{Data viz Combination} column, and on the basis of the 3FRSM analysis, it was possible to choose one combination of items per data viz, such that the corresponding measures lay on a sort of continuum, with increasing measures.
The strategy was, firstly, to select the most discriminating combinations, i.e., B1 and ST4. Secondly, for A, TM and P, we selected the ones with the highest measure not shared with other combinations, i.e., A1, TM1, and P2. We then selected G12 as a G combination, because it was mid-positioned in the discriminability scale, among five other configurations of G.
L3 was then selected among L combinations because, of the two with the higher discriminability power, it was the one with the highest \textit{PtMea Corr}.
Lastly, SC1 was added as the unique SC combination.
Therefore, the final set of items consisted of 17 multiple-choice items, 5 polytomous items, one true/false item and 9 free-text items. The total number of selected items was 32. Table~\ref{tab:items} reports, for each data viz, the number associated with the selected items and the corresponding codes that were used in the subsequent Rasch analysis. 

\begin{table}[htb]
  \caption{List of selected items by data viz and tasks. The first rows report the numbers associated with the items, whereas the second rows the codes used in the Rasch analysis.}
  \label{tab:items}
  \scriptsize%
  \centering%
  \begin{tabular}{llcccc}
    \hline
    Data Viz & &Name &  Represent & Use & Content \\
    \hline
    A &  Item & 37 & 38 & 39 & 40 \\
                      & Code & A & A\_Repr & A\_Use & A\_Cont \\
    B & Item &42  & 43 & 44  & 45\\
                & Code  & B & B\_Repr & B\_Use & B\_Cont \\
    G &Item & 1 & 3 & 5 & 8\\
                  & Code  & G & G\_Repr & G\_Use & G\_Cont \\
    L &Item & 19 & 20 & 21 & 25\\
              & Code  & L & L\_Repr & L\_Use & L\_Cont\\
    P & Item & 32 & 34 & 33  & 36\\
             & Code  & P & P\_Repr & P\_Use & P\_Cont\\
    SC &Item & 47 & 48 & 49 & 50\\
               & Code   & SC & SC\_Repr & SC\_Use & SC\_Cont\\
    ST &Item & 11  & 13 & 15 & 17\\
                & Code      & ST & ST\_Repr & ST\_Use & ST\_Cont\\
    TM &Item & 27 & 28 & 29 & 30\\
           & Code   & TM & TM\_Repr & TM\_Use & TM\_Cont\\
  	\hline
  \end{tabular}%
\end{table}

%\subsection{The questionnaire}

The items identified in the previous step were administered to two high-school students in a private session with one of the authors, for adjusting the wording of the items. The items were then administered in a questionnaire to a sample of 75 high-school students, all attending the 4th or 5th grade classes of an Italian scientific high-school (60), and an Italian technical high-school (15). In particular, the latter students underwent a mini-course about data viz design (10 hours) some weeks before taking the test. The former did not undergo any previous specific course about data visualization, but the professor responsible for their vigilance during the test suggested that they may have some literacy about data viz due to their scientific curricula. The sample was equally divided into females and males. The age of the participant was between 18 and 19 years old. 

In order to analyse the responses with the most suitable model belonging to the family of Rasch models, the 17 multiple-choice items, the true/false item and the 9 free text items were dichotomised, with code 1 for the correct answer and 0 for any other answer. Therefore, the item bank consisted of 27 dychotomous and 5 polytomous (TM\_Use, P\_Repr, P\_Cont, B\_Use and SC\_Use) items. Given the nature of the items involved, comprising both dichotomous and polytomous formats, we employed the PCM to analyse the results, as it allows for the estimation of distinct thresholds across items. The mean of item difficulty estimates was set to 0.0 logits and the estimates of the parameters were obtained using the (unconditional) maximum likelihood estimation method\footnote{All the analyses were performed using the WINSTEP software~\cite{WINSTEP}}. 

The inspection of the thresholds for the five polytomous items revealed that the two thresholds of item SC\_Use  were inverted ($\tau_1=1.50$, $\tau_2=-1.50$), whereas the two thresholds of item B\_Use  were too close ($\tau_1=-0.03$, $\tau_2=0.03$). So, we decided to collapse, for both items, the first two categories, obtaining two dichotomous items. 

As stated before, the successful implementation of PCM necessitates that the assumptions of local independence and unidimensionality are fulfilled, meaning that the items should approximate both local independence and unidimensionality. One tool to detect local independence is the item Pearson correlation, which was computed using residuals across all subjects who responded to both items. Potentially locally dependent pairs of items should have high positive or negative correlations. 

The principal component analysis (PCA) on standardised residuals\footnote{the Rasch method uses the PCA in a counterintuitive, yet valid way: excluding multi-dimensionality in the items.} was used to verify the assumption of unidimensionality. In the Rasch framework, the purpose of conducting this analysis is not to identify shared factors. The underlying hypothesis is that the data contain only one dimension (the Rasch dimension) caught by the model, so the residuals should not contain other significant dimensions. 
Thus, the objective was to confirm the unidimensionality assumption~\cite{BondFox2015}.

The expected statistical outcomes of applying RM analysis to the pilot study data are available in Table~\ref{tab:pilot} in Appendix~\ref{sec:appendix}.

\begin{figure}[htb]
\scriptsize
\centering
\includegraphics[width=\columnwidth]{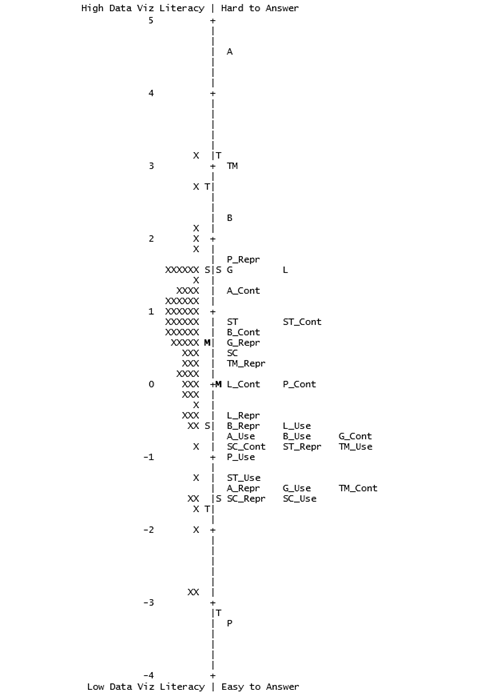}
\caption{Wright map of the PCM analysis of the pilot study. Labels of each item contain only the data viz abbreviation for the \emph{Name} task, abbreviation and \emph{\_Cont, \_Repr, and \_Use} for the other three tasks, respectively.}
\label{fig:WM_pilot}
\end{figure}

The distribution of students' levels of data viz literacy and item difficulties can be simultaneously 
displayed in a single visualization called \textit{Wright map} (Figure~\ref{fig:WM_pilot}), provided that both measures are expressed in logits.
“M” marks the student and item mean, “S” indicates one sample standard deviation from the mean and “T” indicates 
two sample standard deviations from the mean.
The students situated at the upper end of the scale exhibited higher levels of data viz literacy, whereas those at the lower end exhibited lower levels. 
The items at the bottom of the scale were considered easy, meaning that most students could answer them correctly. In contrast, the items at the top of the scale were considered hard, which means that most students were unable to answer them correctly.

\section{Discussion}
\label{sec:discussion}

\subsection{The 3MFRM model}
The application to the original data showed disordered thresholds for higher categories (as reported in Section~\ref{sec:mfrmres}), and this may suggest that the rating scale did not work properly. One interpretation may be that the raters showed some difficulty distinguishing between fairly and very difficult when they rated the data viz items. 

Looking at the separation reliability indices for examinees (i.e. data viz combinations), tasks, and raters, we found that:
\begin{itemize}
\item the value of R for examinees is 0.68, indicating a moderate level of separability, likely due to the specific type of examinees used in this analysis. For example, the stacked area chart yielded two examinees, A1 and A2, which only differ by one item (\textit{Content}); 
\item the value of R for tasks is 0.96. This indicates that the tasks are highly distinct from each other; 
\item the value of R for raters is 0.84, which means that the raters do not agree in their way to judge the difficulty of the tasks. This finding is confirmed by the rejection of the null hypothesis of equal severity of the raters, tested by a Wald-statistic (p-value $<$ 0.001)~\cite{Eckes2015}, and by the presentation of raters' disagreement on each item, reported in Section~\ref{sec:irr}.
\end{itemize}

Concerning the fit statistics, one of the two combinations involving the pie chart, namely P1, presented the highest values of Infit/Outfit MNSQ statistics, though still within the limits of real misfitting. Nevertheless, this finding suggests looking more carefully at P1. Its $PtMea Corr$ is $-0.12$, indicating a departure of its measure from its ability to better discriminate the level of data viz literacy from the model expectation. P1 differs from P2, the other combination that still involves a pie chart, by item 35 (task \textit{Content}). As shown in Figure~\ref{fig:heatmaps1} of Section~\ref{sec:irr}, this item exhibited a high degree of disagreement among raters, mainly R5 and R6. These two raters appeared to be the ones who gave this item the unexpected score of 4, i.e., fairly or very difficult on the original scale, instead of 1.8, i.e., middle easy, as expected by the estimated 3MFRM.

Moreover, taking into account the $PtMea Corr$ index, two other data viz combinations showed negative values: ST8, with -0.07, and ST6, with -0.06. These are two of the eight combinations involving the stacked bar chart. Two combinations, namely ST1 and ST3, had the higher $PtMea Corr$ indices instead (respectively 0.74 and 0.44), and what differentiated them from ST8 and ST6 was the \textit{Represent} task (item 12 for ST1/ST3, and item 13 for ST8/ST6), as well as the \textit{Content} task (item 17 for ST1/ST3, and item 18 for ST8/ST6). Looking at the heatmap in Figure~\ref{fig:heatmaps1}, items 17 and 18 exhibited a similar degree of disagreement, while item 12 showed a slightly higher level of disagreement than item 13. %This may suggest that the item wording in this case could be the main cause of the raters' disagreement.

Inspection of the fit indices for the raters did not reveal any discrepancy between their observed and expected behaviour. The fit statistics for the tasks highlighted some criticality for task \textit{Name}, which exhibited high values of Infit and Outfit MNSQ statistics (Infit MNSQ = 1.40; Outfit MNSQ = 1.41), although not high enough to indicate misfitting. This finding may suggest that, although the task \textit{Name} is related to the data viz literacy construct, its connection appears weaker than those shown by the other three tasks.

In order to investigate whether the judgment process was fair and whether each rater kept a uniform level of severity among the examinees, we tested the significance of the examinee-by-rate interaction parameters $\phi_{nj}$ in formula~\ref{eq:MFRM_int}. Our analysis did not reveal any significant parameter, allowing us to conclude that each rater maintained a uniform level of severity across the data viz combinations.

Regarding the Wright Map, bubble chart (B\_ combinations) emerged as the data viz with the highest discriminability effect, with both B1 and B2 receiving the highest scores on the difficulty scale. On the other hand, pie chart (P\_ combinations) appeared as the data viz that discriminated the least, given that both P1 and P2 received lower scores on the difficulty scale. This result may not be surprising; the pie chart is commonly taught in primary education level curricula, thus the kind of information that can be read from it may be the most intuitive. It is reasonable that the raters judged its tasks very easy. 

Most of the 8 combinations involving choropleth map (G\_) appeared in the lower part of the map, suggesting that this data viz followed the easiest one. Choropleth map is a popular data viz, fairly easy to interpret and widely used for visualising geospatial data across various media, including the news.
 
Two combinations of L, namely L3 and L4, had a higher than average discriminability effect (-0.19), while the other two combinations, namely L1 and L2, were below this average. This kind of result may be useful when assembling a new test. 

Indeed, the inspection of the 3MFRM model allowed us to infer that both data viz discriminability and representativeness may conflict; thus, DRIVE-T may help make a decision based on a justified trade-off between the two. For example, other combinations can be selected before L, and the best discriminating L combination can be added in the end. By applying DRIVE-T, the hierarchical progression of data viz and tasks representativeness is improved, unwanted interactions are avoided, and the resulting test may be more reliable and effective. 

%SPOSTATO NEI RESULTS DEL PILOT STUDY
%For example, for identifying the more reliable and effective version of our test to be used in the pilot study reported in Section~\ref{sec:pilot}, other combinations were selected before the line chart, and the best discrimitating line chart combination was added in the end.  %fill selected to be placed in the selection of all the other combinations, and this could allow the creation of a more reliable and effective final test.

Regarding raters, more severe ones appeared in higher positions, and less severe ones in lower positions. For identifiability reasons, the average measure of raters' severity was constrained to be zero. Raters R2, R5 and R6 had a mean severity. R3 and R4 were the most severe raters, whereas R1 and R7 were the most lenient ones. The variability across raters was low, their measures showing a 0.68-logit spread. %(S.D. = 0.26) 
This may suggest that, even if there was no agreement among raters, as discussed previously, their level of severity was quite similar.
From the visual inspection of the heatmaps in Section~\ref{sec:irr}, it is observable that the task causing more disagreement among raters is the one related to knowing the \textit{Name} of the data viz presented. The task related to retrieving the \textit{Content} of a data viz is showing disagreements among raters, too. The difficulty of tasks related to what a data viz \textit{Represent}(s) and how a data viz is \textit{Use}(d) are mostly agreed by all the raters. These results reflected some patterns: \textit{Name} and \textit{Content} tasks frequently produced disagreements, whereas \textit{Use} and, to some extent, \textit{Represent} had higher consistency. Raters emerged as isolated contributors to disagreement, highlighting individual interpretive differences among them. However, those of the heatmaps were only providing an explorative analysis that further supports the 3MFRM results. 

Regarding tasks, \textit{Name} emerged as the most challenging task, followed by \textit{Content}, \textit{Use} and \textit{Represent}. This hierarchy of measures may suggest the presence of a progression level for the tasks being measured, and this hierarchy may help outline difficulty levels of an underlying construct for data viz literacy.

\subsection{The pilot study}

We note that, according to Figure~\ref{fig:WM} and the results reported in Section~\ref{sec:mfrmres}, the item combination selected among the ones modelled with the 3MFRM, allowed the identification of a hierarchical progression expression of the data viz along the literacy continuum that was better reflected in the test designed to measure it.

The Rasch model used to analyse the pilot study data introduced in Section~\ref{sec:PCM}, yielded a person reliability index of 0.79, sufficiently high to ensure that the questionnaire was sensitive enough to distinguish between students with high and low levels of data viz literacy. The item reliability index was 0.95, sufficiently high to ensure that the sample of students was able to confirm the item difficulty hierarchy of the questionnaire.

The highest positive Pearson correlation value was 0.37 and involved items ST and SC, whereas the highest negative Pearson correlation value was -0.41 and involved the items L\_Repr and TM. As the correlation values were relatively small, and no meaningful reasons justify connections among these items, it can be concluded that the items were approximately locally independent.

Regarding the PCA, the eigenvalues of the first two dimensions were 3.03 and 2.74 respectively, suggesting the possible presence of multidimensionality. Thus, a further analysis was necessary to understand whether the degree of multidimensionality present in the data was substantial enough to justify partitioning the items into separate tests. In order to investigate this issue, for each dimension, the items were clustered into three clusters according to high, low, and middle values of their factor loadings~\cite{WINSTEP_Manual}. Usually, the middle cluster (2) roughly corresponds to the Rasch dimension. Thus, it was important to verify that the correlation between this cluster and the other two was high. Clusters 1 and 3 were both somewhat off-dimension, making the correlation between them somewhat incidental~\cite{WINSTEP_Manual}. In order to account for the measurement error, the disattenuated correlation was used. The disattenuated correlation values for the 2-1 and 2-3 clusters were, respectively, 1 and 0.63 for the first dimension and 0.62 and 0.88 for the second dimension. These values were sufficiently high to indicate that the clusters of items were measuring the same thing.

Looking at the item fit statistics, only item P\_Repr showed a high value for Infit (1.41) and Outfit (2.73) MNSQ. This means that this item had criticalities and deserved a rewriting or a substitution.

Regarding the analysis of the Write Map, the easiest item was the one that asked students to indicate the name of a pie chart. As also emerged from the 3MFRM analysis, given that the pie chart is commonly taught in primary education level curricula, this finding was not surprising. 
Five out of the eight items asking for the name of the data viz were located to the very upper part of the item scale, indicating that the knowledge of the name of a data viz was, in most cases, a difficult task. This result is in line with the findings from the analysis of the raters’ evaluations, where the \textit{Name} task was judged to be the most difficult one.

The average level of data viz literacy observed in the sample of students was 0.51 (SD 1.1), which was greater than the average difficulty of the items (being it zero logits). This discrepancy may suggest that this group of students, on the whole, demonstrated an adequate level of data viz literacy.

\subsection{A comparison between raters and respondents on a common scale}

\begin{figure}[ht]
\centering
\includegraphics[width=.96\columnwidth]{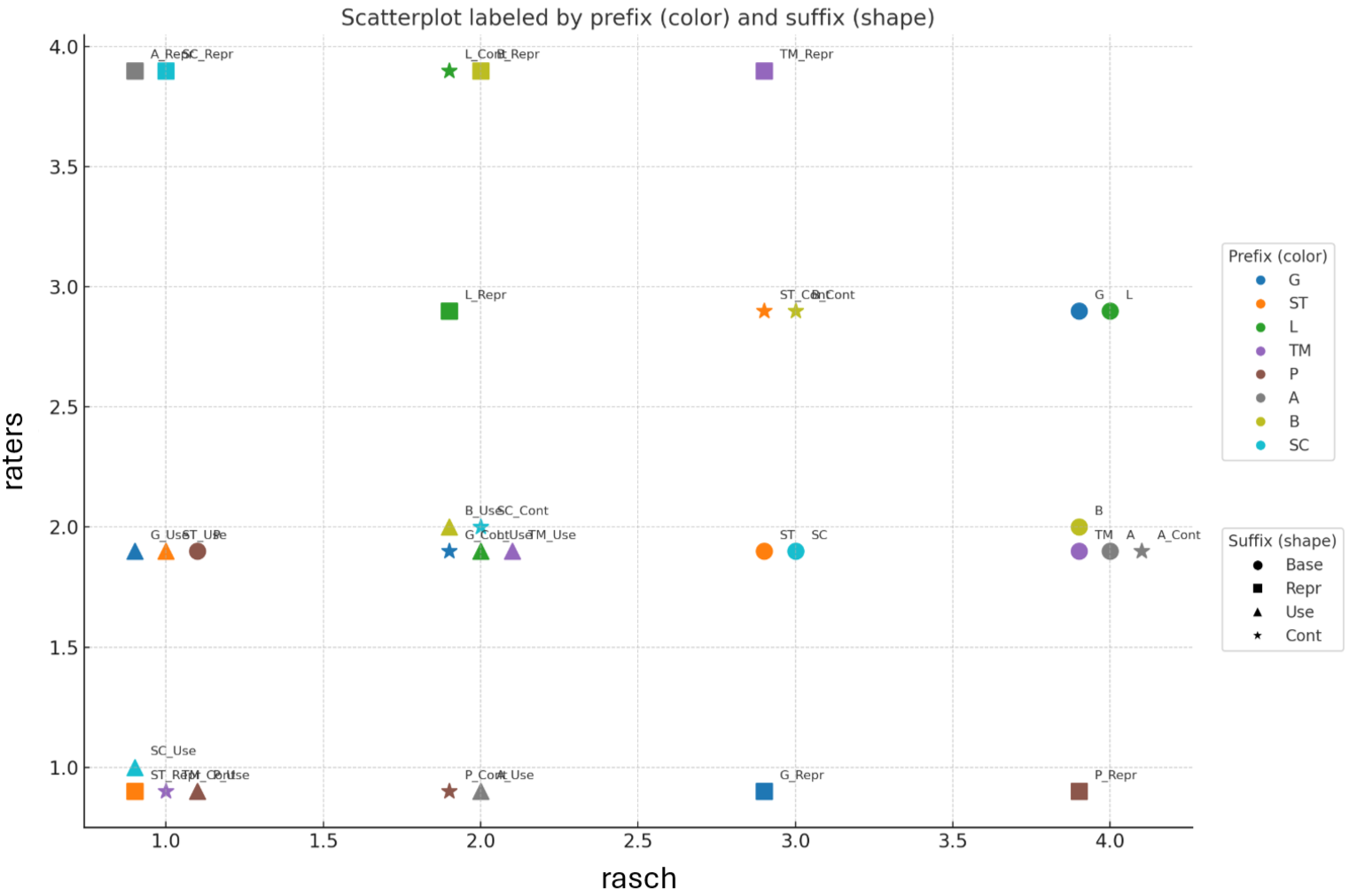}
\caption{The scatterplots putting in correlation raters' scores and pilot study Rasch measures into a categorical classification where 1=very easy, 2=easy, 3=hard, and 4=very hard. Each symbol cluster shares identical coordinates, but was visually offset to enhance
readability.}
\label{fig:ratersrasch}
\end{figure}

Figure~\ref{fig:ratersrasch} displays a scatterplot comparing the raters' median score on each item ($y$ axis) against the Rasch logit score of the pilot study ($x$ axis), after mapping both into a classification scale, where: 1=very easy, 2=easy, 3=hard, and 4=very hard. These mappings consider Rasch measures centered around zero, and took the median of both halves (below and above 0) as further thresholds dividing the set of measures into the 4 categories. Raters' scores were used to map the other dimension of the scatterplot, where the initial median scores of 1 and 2 were assigned category 1, score 3 was assigned category 2, score 4 was assigned category 3, and scores 5 and 6 were assigned category 4. 
Looking at the figure, we can observe what follows for the four tasks: 

\begin{itemize}
\item the triangles are those representing the \textit{Use} tasks, for which both raters and respondents seem to agree in positioning them at the bottom left corner, as the easiest ones, almost independent from the data viz;
\item the squares represent the \textit{Represent} tasks, and they seem to be the most discordant according to where raters and respondents positioned them;
\item the stars represent the \textit{Content} tasks, which seem to lay on a growing linear trajectory, where both raters and respondents put them in concordance, and difficulty levels change depending on the data viz; 
\item the circles represent the \textit{Name} task, showing a concordance between raters and respondents in positioning them at the mid right of the scatterplot, suggesting that both found these items quite difficult, independent of the data viz
\end{itemize}

This comparison may be useful as a double-check analysis with regard to the quality of the assessment test and the target population under consideration.

\subsection{Limitations}
\label{sec:limitations}

Limitations of the work could be raised regarding the number of raters involved in the activity of scoring items. We recall that the main purpose of the paper was to introduce the DRIVE-T methodology, whose intent was to provide a quick and cheap way to qualitatively identify the more discriminative yet task-representative items among a set of ready-to-use or created-from-scratch items for each data viz. For this reason, the methodology should be proved robust enough to be applied with few raters, who were experts in the field. The activity of raters is analogous to what is done during heuristics evaluation approaches, where a panel of experts, usually three to five people\cite{nielsen1994theory}, are asked to evaluate qualitatively the usability of application interfaces within their realm of expertise. Critiques may be moved to the number of items exploited in this study. Again, what was evaluated through the methodology was not the numerosity of items, but rather their expressivity in terms of cognitive traits related to the ability to master the data viz language. For this reason, the items should be qualitatively discriminating and representing all the traits of the measurement construct, as well as the majority of data viz difficulty levels. To reach this goal, it was not the quantity, but rather the quality of items that was prioritized. Another limitation may lay in the use of eight data viz, compared with the wider spectrum of available alternatives. As remarked in the introduction, the complexity of the measurement construct and of the item modelling required to focus on fewer, yet mostly used and qualitatively robust data viz. Relying on a set of them that is representative and common in data viz literacy assessment tests also guarantees future comparability with the other approaches. 
%Finally, an objection may be raised regarding the use of a subset of items for the pilot study, instead of all the items. 

\section{Conclusions}
\label{sec:conc}

In this paper, we showed the potential and limitations of a methodology for the unprecedented rapid prototyping of items expressing latent traits of a data viz literacy measurement construct explicitly hypothesised but still blurred, and before administering items to respondents. DRIVE-T was able to better outline such construct (RQ1), to let emerge constructs from practice, and provide the best potentials of items, before administering the items to respondents (RQ2). Furthermore, modelling the raters' scoring with a 3MFRM, we were able to select the best items (under the criteria of discriminability and representativeness) to create an assessment test (RQ3). This test was then administered to a sample in a pilot study. This pilot study confirmed the quality of the items being designed, rated, and selected, proving their discriminability and representativeness in measuring levels of data viz literacy. %The outlined construct is mostly unidimensional, with the emergence of a second dimension related to items asking the name of the data viz. This interesting result seems to outline a second dimension of the hypothesised construct. 
Furthermore, all the tasks expressing latent traits of the data viz literacy construct identified with DRIVE-T and analyzed in the space of a Rasch-based pilot study have been compared on a common scale in search of concordant and discordant patterns between raters and respondents. 

This work paved the way for an unprecedented method for measurement construct design, and the identification of its hierarchical and progressive levels, with a minimum effort (time and cost) towards high-quality tests.

Constructs-foregrounding is possible when the most discriminative and representative items are selected, which may help the community of test designers scrutinise, discuss and adjust items when they want to design or make sense of existing data visualization literacy assessment tests, whenever no data visualization literacy measurement construct was explicitly or completely outlined beforehand.

%\textcolor{red}{As for P, also L exhibited a high disagreement among raters, especially for the \textit{Content} task, and both were estimated as being zero logits in the pilot study Write Map.}
%Interestingly, \textit{Content} items were highly disagreed by raters with respect to other task-based items. This seems to be also reflected on the Write Map of the pilot study, where the majority of \textit{Content} items are equal or above zero logits.}

%\subsection{Construct outline}
%What do the model tell us about the construct? Levels in order of difficulty (this may justify items of a certain kind other than dichotomous)...

%% if specified like this the section will be omitted in review mode
%% anonimizzato - RIPRISTINARE!!!
\section*{Acknowledgements}
% The authors wish to thank Sara Beschi, Manuel Mastrofini, Marco Sandri and Matteo Ventura.
%This work was supported by the Research funded by the European Union – Next-Generation, Mission 4 Component 1 CUP D53D23008690006, project name: ``Characterizing and Measuring Visual Information Literacy'' ID 2022JJ3PA5. 

\bibliographystyle{IEEEtran} 
\bibliography{template}

\begin{thebibliography}{10}
\providecommand{\url}[1]{#1}
\csname url@samestyle\endcsname
\providecommand{\newblock}{\relax}
\providecommand{\bibinfo}[2]{#2}
\providecommand{\BIBentrySTDinterwordspacing}{\spaceskip=0pt\relax}
\providecommand{\BIBentryALTinterwordstretchfactor}{4}
\providecommand{\BIBentryALTinterwordspacing}{\spaceskip=\fontdimen2\font plus
\BIBentryALTinterwordstretchfactor\fontdimen3\font minus \fontdimen4\font\relax}
\providecommand{\BIBforeignlanguage}[2]{{%
\expandafter\ifx\csname l@#1\endcsname\relax
\typeout{** WARNING: IEEEtran.bst: No hyphenation pattern has been}%
\typeout{** loaded for the language `#1'. Using the pattern for}%
\typeout{** the default language instead.}%
\else
\language=\csname l@#1\endcsname
\fi
#2}}
\providecommand{\BIBdecl}{\relax}
\BIBdecl

\bibitem{aoyama2003graph}
K.~Aoyama and M.~Stephens, ``Graph interpretation aspects of statistical literacy: A japanese perspective,'' \emph{Mathematics Education Research Journal}, vol.~15, no.~3, pp. 207--225, 2003.

\bibitem{Galesicetal2011}
M.~Galesic and R.~Garcia-Retamero, ``Graph literacy: A cross-cultural comparison,'' \emph{Medical Decision Making}, vol.~31, no.~3, pp. 444--457, 2011.

\bibitem{Merbitzetal1989}
C.~Merbitz, J.~Morris, and J.~C. Grip, ``Ordinal scales and foundations of misinference,'' \emph{Archives of Physical Medicine and Rehabilitation}, vol.~70, no.~4, pp. 308--312, 1989.

\bibitem{Lee2017551}
S.~Lee, S.-H. Kim, and B.~C. Kwon, ``Vlat: Development of a visualization literacy assessment test,'' \emph{IEEE Transactions on Visualization and Computer Graphics}, vol.~23, no.~1, p. 551 – 560, 2017.

\bibitem{krejci2020visual}
S.~E. Krejci, S.~Ramroop-Butts, H.~N. Torres, and R.~D. Isokpehi, ``Visual literacy intervention for improving undergraduate student critical thinking of global sustainability issues,'' \emph{Sustainability}, vol.~12, no.~23, 2020.

\bibitem{locoro2021visual}
A.~Locoro, W.~P. Fisher, and L.~Mari, ``Visual information literacy: Definition, construct modeling and assessment,'' \emph{IEEE access}, vol.~9, pp. 71\,053--71\,071, 2021.

\bibitem{yang2021explaining}
L.~Yang, C.~Xiong, J.~K. Wong, A.~Wu, and H.~Qu, ``Explaining with examples: Lessons learned from crowdsourced introductory description of information visualizations,'' \emph{IEEE Transactions on Visualization and Computer Graphics}, vol.~29, no.~3, pp. 1638--1650, 2021.

\bibitem{camba2022identifying}
J.~D. Camba, P.~Company, and V.~L. Byrd, ``Identifying deception as a critical component of visualization literacy,'' \emph{IEEE Computer Graphics and Applications}, vol.~42, no.~1, pp. 116--122, 2022.

\bibitem{firat2022p}
E.~E. Firat, A.~Denisova, M.~L. Wilson, and R.~S. Laramee, ``P-lite: A study of parallel coordinate plot literacy,'' \emph{Visual Informatics}, vol.~6, no.~3, pp. 81--99, 2022.

\bibitem{CALVI}
L.~W. Ge, Y.~Cui, and M.~Kay, ``Calvi: Critical thinking assessment for literacy in visualizations,'' in \emph{Proceedings of the 2023 CHI Conference on Human Factors in Computing Systems}, ser. CHI '23.\hskip 1em plus 0.5em minus 0.4em\relax New York, NY, USA: Association for Computing Machinery, 2023, pp. 1--18.

\bibitem{davis2024risks}
R.~Davis, X.~Pu, Y.~Ding, B.~D. Hall, K.~Bonilla, M.~Feng, M.~Kay, and L.~Harrison, ``The risks of ranking: Revisiting graphical perception to model individual differences in visualization performance,'' \emph{IEEE Transactions on Visualization and Computer Graphics}, vol.~30, no.~3, pp. 1756--1771, 2024.

\bibitem{Cui-Promises-Pitfalls}
Y.~Cui, L.~W. Ge, Y.~Ding, L.~Harrison, F.~Yang, and M.~Kay, ``Promises and pitfalls: Using large language models to generate visualization items,'' \emph{IEEE Transactions on Visualization and Computer Graphics}, pp. 1--11, 2024.

\bibitem{bloom1956taxonomy}
B.~S. Bloom \emph{et~al.}, ``Taxonomy of educational objectives. vol. 1: Cognitive domain,'' \emph{New York: McKay}, vol.~20, p.~24, 1956.

\bibitem{pinker1990theory}
S.~Pinker, ``A theory of graph comprehension,'' \emph{Artificial intelligence and the future of testing}, pp. 73--126, 1990.

\bibitem{friel2001making}
S.~N. Friel, F.~R. Curcio, and G.~W. Bright, ``Making sense of graphs: Critical factors influencing comprehension and instructional implications,'' \emph{Journal for Research in mathematics Education}, vol.~32, no.~2, pp. 124--158, 2001.

\bibitem{burns2020evaluate}
A.~Burns, C.~Xiong, S.~Franconeri, A.~Cairo, and N.~Mahyar, ``How to evaluate data visualizations across different levels of understanding,'' \emph{arXiv preprint arXiv:2009.01747}, 2020.

\bibitem{boy2014principled}
J.~Boy, R.~A. Rensink, E.~Bertini, and J.-D. Fekete, ``A principled way of assessing visualization literacy,'' \emph{IEEE Transactions on Visualization and Computer Graphics}, vol.~20, no.~12, pp. 1963--1972, 2014.

\bibitem{Beschietal2025}
S.~Beschi, D.~Falessi, S.~Golia, and A.~Locoro, ``Characterizing data visualization literacy for standardization: A systematic literature review,'' \emph{IEEE Access}, vol.~13, pp. 65\,704--65\,725, 2025.

\bibitem{mari2023measurement}
L.~Mari, M.~Wilson, and A.~Maul, \emph{Measurement across the sciences: Developing a shared concept system for measurement}.\hskip 1em plus 0.5em minus 0.4em\relax Springer Nature, 2023.

\bibitem{Pandey20231}
S.~Pandey and A.~Ottley, ``Mini-vlat: A short and effective measure of visualization literacy,'' \emph{Computer Graphics Forum}, vol.~42, no.~3, p. 1 – 11, 2023.

\bibitem{Okanetal2019}
Y.~Okan, E.~Janssen, M.~Galesic, and E.~A. Waters, ``Using the short graph literacy scale to predict precursors of health behavior change,'' \emph{Medical Decision Making}, vol.~39, no.~3, pp. 183--195, 2019.

\bibitem{Cabitza2014MadeWK}
\BIBentryALTinterwordspacing
F.~Cabitza and A.~Locoro, ``"made with knowledge" - disentangling the it knowledge artifact by a qualitative literature review,'' in \emph{International Joint Conference on Knowledge Discovery, Knowledge Engineering and Knowledge Management}, 2014, pp. 64--75. [Online]. Available: \url{https://api.semanticscholar.org/CorpusID:28985187}
\BIBentrySTDinterwordspacing

\bibitem{wilson2004constructing}
M.~Wilson, \emph{Constructing measures: An item response modeling approach}.\hskip 1em plus 0.5em minus 0.4em\relax Routledge, 2004.

\bibitem{aristotle2006nicomachean}
.~Aristotle, \emph{Nicomachean ethics}.\hskip 1em plus 0.5em minus 0.4em\relax ReadHowYouWant. com, 2006.

\bibitem{peirce1931semiotics}
C.~S. Peirce, \emph{Semiotics}.\hskip 1em plus 0.5em minus 0.4em\relax Translation Jos{\'e} Teixeira Coelho Netto, 1931, vol.~8.

\bibitem{10756169}
J.~T. Otto and S.~Davidoff, ``Visualization artifacts are boundary objects,'' in \emph{2024 IEEE Evaluation and Beyond - Methodological Approaches for Visualization (BELIV)}, 2024, pp. 81--88.

\bibitem{de2005semiotic}
C.~S. De~Souza, \emph{The semiotic engineering of human-computer interaction}.\hskip 1em plus 0.5em minus 0.4em\relax MIT press, 2005.

\bibitem{shneiderman2003eyes}
B.~Shneiderman, ``The eyes have it: A task by data type taxonomy for information visualizations,'' in \emph{The craft of information visualization}.\hskip 1em plus 0.5em minus 0.4em\relax Elsevier, 2003, pp. 364--371.

\bibitem{Eckes2015}
T.~Eckes, \emph{Introduction to Many-Facet Rasch Measurement: Analyzing and Evaluating Rater-Mediated Assessments. 2nd Revised and Updated Edition}.\hskip 1em plus 0.5em minus 0.4em\relax Peter Lang Edition, 2015.

\bibitem{marasini2016assessing}
D.~Marasini, P.~Quatto, and E.~Ripamonti, ``Assessing the inter-rater agreement for ordinal data through weighted indexes,'' \emph{Statistical methods in medical research}, vol.~25, no.~6, pp. 2611--2633, 2016.

\bibitem{andrich1988rasch}
D.~Andrich, \emph{Rasch models for measurement: Sage publications}.\hskip 1em plus 0.5em minus 0.4em\relax Sage Publications, 1988.

\bibitem{Rasch1960}
G.~Rasch, \emph{Probabilistic models for some intelligence and attainment tests}.\hskip 1em plus 0.5em minus 0.4em\relax Danish Institute for Educational Research. (Expanded edition, 1980. Chicago: University of Chicago Press.), 1960.

\bibitem{Andrich1978}
D.~Andrich, ``A rating formulation for ordered response categories,'' \emph{Psychometrika}, vol.~43, no.~4, pp. 561--573, 1978.

\bibitem{Masters1982}
G.~N. Masters, ``A rasch model for partial credit scoring,'' \emph{Psychometrika}, vol.~47, no.~2, pp. 149--174, 1982.

\bibitem{Linacre1989}
J.~M. Linacre, \emph{Many-facet Rasch measurement}.\hskip 1em plus 0.5em minus 0.4em\relax Chicago: MESA Press, 1989.

\bibitem{Linacre2002}
------, ``What do infit and outfit, mean-square and standardized means?'' \emph{Rasch Measurement Transactions}, vol.~16, p. 878, 2002.

\bibitem{MyfordWolfe2003}
C.~Myford and E.~W. Wolfe, ``Detecting and measuring rater effects using many-facet rasch measurement: Part i,'' \emph{Journal of Applied Measurement}, vol.~4, no.~4, pp. 386--422, 2003.

\bibitem{Eckes2009}
T.~Eckes, ``Many-facet rasch measurement,'' in \emph{Reference supplement to the Manual for relating language examinations to the Common European Framework of Reference for languages: Learning teaching, assessment}.\hskip 1em plus 0.5em minus 0.4em\relax Council of Europe/Language Policy Division, 2009, p. Section H.

\bibitem{LinacreWrite2002}
J.~M. Linacre and B.~D. Wright, ``Construction of measures from many-facet data,'' \emph{Journal of Applied Measurement}, vol.~3, p. 484–509, 2002.

\bibitem{FACET}
\BIBentryALTinterwordspacing
J.~M. Linacre, ``Facets computer program for many-facet rasch measurement,'' 2025. [Online]. Available: \url{https://www.winsteps.com/facets.htm}
\BIBentrySTDinterwordspacing

\bibitem{WINSTEP}
\BIBentryALTinterwordspacing
------, ``Winsteps\textsuperscript{\textregistered} rasch measurement computer program (version 5.4.0),'' 2023. [Online]. Available: \url{https://www.winsteps.com}
\BIBentrySTDinterwordspacing

\bibitem{BondFox2015}
T.~G. Bond and C.~M. Fox, \emph{Applying the Rasch Model: Fundamental Measurement in the Human Sciences, Third Edition}.\hskip 1em plus 0.5em minus 0.4em\relax New York: Routledge, 2015.

\bibitem{WINSTEP_Manual}
J.~M. Linacre, \emph{Winsteps\textsuperscript{\textregistered} Rasch measurement computer program User's Guide. Version 5.4.0}.\hskip 1em plus 0.5em minus 0.4em\relax Portland, Oregon: Winsteps.com, 2023.

\bibitem{nielsen1994theory}
J.~Nielsen, ``The theory behind heuristic evaluations,'' \emph{Nielsen Normal Group}, 1994.

\end{thebibliography}

\newpage

%\begin{IEEEbiography}[{\includegraphics[width=1in,height=1in,clip,keepaspectratio]{author4.jpeg}}]{Angela Locoro} is an Associate Professor of Artificial Intelligence at the University of Brescia, Italy. She is an Editorial Board Member of the Information Fusion Journal. Her main research interests are in the DV field, where she was funded for a project on DVL. She received her Ph.D. in Computer Engineering, MA in Modern Literature, and BSc in Computer Science from the University of Genova. 
%\end{IEEEbiography}

%\begin{IEEEbiography}[{\includegraphics[width=1in,height=1in,clip,keepaspectratio]{author3.png}}]{Silvia Golia} is an Assistant Professor of Statistics at the University of Brescia, Italy. She received her MSc Degree in Statistics from the University of Padua and her PhD in Statistics from the University of Perugia. Her research interests include Rasch Analysis, Causal Analysis, Bayesian Networks, Decision Trees and Time series Analysis. \end{IEEEbiography}

%\begin{IEEEbiography}[{\includegraphics[width=1in,height=1in,clip,keepaspectratio]{author2.jpg}}]{Davide Falessi} is an Associate Professor of Software Engineering at the University of Rome Tor Vergata, Italy. He is the Associate Editor in Software Economics of IEEE Software and an Editorial Board member of the Empirical Software Engineering Journal.  His main research interest is in devising and empirically assessing scalable solutions for developing software-intensive systems. He received his Ph.D., MSc, and BSc degrees in Computer Engineering from the University of Rome Tor Vergata, Italy.\end{IEEEbiography}

\appendix
\section{}
\label{sec:appendix}

\begin{table}[ht]
\scriptsize
\centering
\caption{Discrimination abilities and fit statistics of the data viz combinations in the 3MFRM Model.} 
\label{tab:mainq}
\begin{tabular}{lcccc}
  \hline
Data viz combination & Measure & Infit MNSQ & Outfit MNSQ & PtMea Corr \\ 
  \hline
B1 & 0.52 & 0.93 & 1.02 & 0.18 \\ 
  B2 & 0.40 & 1.18 & 1.22 & 0.22 \\ 
  ST4 & 0.25 & 1.09 & 1.19 & 0.02 \\ 
  ST8 & 0.18 & 1.26 & 1.35 & -0.07 \\ 
  L4 & 0.14 & 0.94 & 0.91 & 0.31 \\ 
  L3 & 0.10 & 0.81 & 0.79 & 0.40 \\ 
  ST2 & 0.07 & 1.04 & 1.09 & 0.04 \\ 
  A1 & 0.03 & 0.77 & 0.77 & 0.65 \\ 
  ST6 & -0.00 & 1.19 & 1.24 & -0.06 \\ 
  TM2 & -0.00 & 1.04 & 1.04 & 0.64 \\ 
  A2 & -0.04 & 0.79 & 0.78 & 0.63 \\ 
  ST3 & -0.11 & 1.08 & 1.05 & 0.44 \\ 
  TM1 & -0.11 & 1.08 & 1.07 & 0.60 \\ 
  G7 & -0.18 & 0.80 & 0.79 & 0.53 \\ 
  SC1 & -0.18 & 1.24 & 1.19 & 0.55 \\ 
  ST1 & -0.18 & 1.04 & 1.01 & 0.74 \\ 
  ST7 & -0.18 & 1.23 & 1.18 & 0.35 \\ 
  G8 & -0.22 & 0.76 & 0.73 & 0.58 \\ 
  L2 & -0.29 & 1.00 & 1.01 & 0.20 \\ 
  L1 & -0.33 & 0.85 & 0.83 & 0.28 \\ 
  G5 & -0.37 & 0.80 & 0.79 & 0.55 \\ 
  ST5 & -0.37 & 1.09 & 1.03 & 0.39 \\ 
  G6 & -0.41 & 0.75 & 0.71 & 0.60 \\ 
  G12 & -0.52 & 1.04 & 1.00 & 0.44 \\ 
  G10 & -0.57 & 0.93 & 0.91 & 0.39 \\ 
  G11 & -0.61 & 0.87 & 0.81 & 0.44 \\ 
  P2 & -0.69 & 1.16 & 1.26 & 0.01 \\ 
  G9 & -0.83 & 0.77 & 0.71 & 0.50 \\ 
  P1 & -0.88 & 1.51 & 1.67 & -0.12 \\ 
   \hline
\end{tabular}
\end{table}

\begin{table}[t]
\scriptsize
\centering
\caption{Task-based items difficulty and their fit statistics in the Rasch analysis of the pilot study.} 
\label{tab:pilot}
\begin{tabular}{lccc}
  \hline
Item & Measure & Infit MNSQ & Outfit MNSQ  \\ 
  \hline
A & 4.54 & 0.86 & 0.35 \\ 
  TM & 2.97 & 1.00 & 0.66 \\ 
  B & 2.23 & 0.82 & 0.63 \\ 
  P\_Repr & 1.74 & 1.41 & 2.73 \\ 
  L & 1.60 & 1.03 & 1.10 \\ 
  G & 1.53 & 1.04 & 1.34 \\ 
  A\_Cont & 1.28 & 1.17 & 1.48 \\ 
  ST & 0.87 & 1.00 & 0.94 \\ 
  ST\_Cont & 0.87 & 1.07 & 1.08 \\ 
  B\_Cont & 0.65 & 1.18 & 1.34 \\ 
  G\_Repr & 0.56 & 1.03 & 0.96 \\ 
  SC & 0.42 & 0.89 & 0.84 \\ 
  TM\_Repr & 0.28 & 1.00 & 0.96 \\ 
  P\_Cont & -0.04 & 1.25 & 1.24 \\ 
  L\_Cont & -0.07 & 0.88 & 0.81 \\ 
  L\_Repr & -0.44 & 1.03 & 1.09 \\ 
  L\_Use & -0.52 & 1.24 & 1.27 \\ 
  B\_Repr & -0.57 & 1.10 & 1.20 \\ 
  B\_Use & -0.74 & 1.00 & 1.03 \\ 
  A\_Use & -0.76 & 0.91 & 0.86 \\ 
  G\_Cont & -0.77 & 0.69 & 0.55 \\ 
  SC\_Cont & -0.85 & 0.88 & 0.80 \\ 
  TM\_Use & -0.87 & 1.00 & 0.97 \\ 
  ST\_Repr & -0.91 & 0.89 & 0.82 \\ 
  P\_Use & -0.93 & 0.86 & 0.74 \\ 
  ST\_Use & -1.31 & 0.71 & 0.56 \\ 
  TM\_Cont & -1.36 & 0.65 & 0.52 \\ 
  G\_Use & -1.41 & 1.05 & 0.85 \\ 
  A\_Repr & -1.48 & 0.91 & 0.74 \\ 
  SC\_Repr & -1.57 & 1.03 & 0.84 \\ 
  SC\_Use & -1.61 & 0.79 & 0.70 \\ 
  P & -3.31 & 0.93 & 0.48 \\ 
   \hline
\end{tabular}
\end{table}

\end{document}